\documentclass[aps,prx,reprint,superscriptaddress,amsmath,amssymb]{revtex4-1}

\usepackage[pdftex]{graphicx}
\usepackage[pdftex,
            colorlinks=true,
            citecolor=blue,
            urlcolor=blue,
            linkcolor=blue]{hyperref}
\usepackage{newtxtext, newtxmath}
\usepackage{bm}
\usepackage{bbm}
\usepackage{braket, mathtools}
\usepackage{tabularx, multirow}
\newcommand{\average}[1]{\ensuremath{\langle#1\rangle} }

\begin{document}

\title{Anapole superconductivity from $\mathcal{PT}$-symmetric mixed-parity interband pairing}

\author{Shota Kanasugi}
 \email{kanasugi.shouta.62w@st.kyoto-u.ac.jp}
 \affiliation{%
 Department of Physics, Kyoto University, Kyoto 606-8502, Japan
}%
\author{Youichi Yanase}%
\email{yanase@scphys.kyoto-u.ac.jp}
\affiliation{%
 Department of Physics, Kyoto University, Kyoto 606-8502, Japan
}%
\affiliation{%
 Institute for Molecular Science, Okazaki 444-8585, Japan
}%

\date{\today}

\begin{abstract}
Recently, superconductivity with spontaneous time-reversal or parity symmetry breaking is attracting much attention owing to its exotic properties, such as nontrivial topology and nonreciprocal transport. 
Particularly fascinating phenomena are expected when the time-reversal and parity symmetry are simultaneously broken. 
This work shows that time-reversal symmetry-breaking mixed-parity superconducting states generally exhibit an unusual asymmetric Bogoliubov spectrum due to nonunitary interband pairing. 
For generic two-band models, we derive the necessary conditions for the asymmetric Bogoliubov spectrum. 
We also demonstrate that the asymmetric Bogoliubov quasiparticles lead to the effective anapole moment of the superconducting state, which stabilizes a nonuniform Fulde-Ferrell-Larkin-Ovchinnikov state at zero magnetic fields. 
The concept of anapole order employed in nuclear physics, magnetic materials science, strongly correlated electron systems, and optoelectronics is extended to superconductors by this work. 
Our conclusions are relevant for any multiband superconductors with competing even- and odd-parity pairing channels. 
Especially, we discuss the superconductivity in UTe$_2$. 
\end{abstract}

\maketitle

\section*{Introduction}
Parity symmetry ($\mathcal{P}$-symmetry) and time-reversal symmetry ($\mathcal{T}$-symmetry) are fundamental properties of quantum materials, such as insulators, metals, magnets, and superconductors. Superconductivity is caused by the quantum condensation of either even-parity or odd-parity Cooper pairs corresponding to spin-singlet or spin-triplet superconductivity due to the fermion antisymmetry~\cite{Sigrist-Ueda}. The order parameter of conventional superconductors breaks neither $\mathcal{P}$-symmetry nor $\mathcal{T}$-symmetry. However, 
competition and coexistence of multiple pairing instabilities lead to exotic superconductivity, such as chiral superconductivity with spontaneous $\mathcal{T}$-symmetry breaking~\cite{Leggett1975} related to the nontrivial topology~\cite{Hasan2010-topo,Qi2011-topo} and anomalous transport~\cite{Taylor2012AHE}. 

\begin{figure*}[htbp]
\centering
   \includegraphics[width=\linewidth]{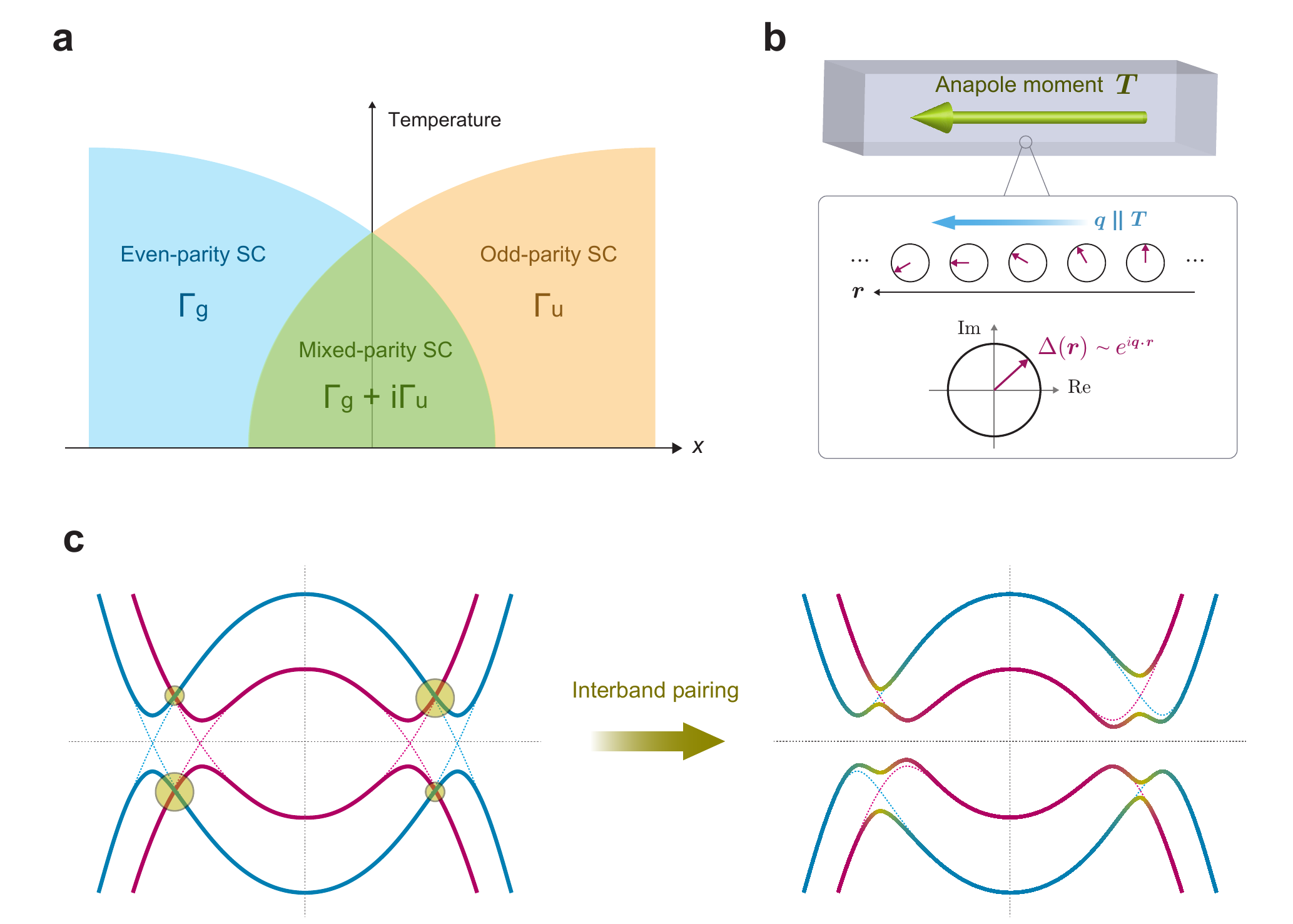}
\caption{\textbf{Schematics of $\mathcal{PT}$-symmetric mixed-parity superconductivity.} 
(a) Schematic phase diagram in a superconductor with comparable strength of even- and odd-parity pairing interactions. The transition between the even-parity superconducting phase ($\Gamma_g$) and the odd-parity superconducting phase ($\Gamma_u$) is induced by tuning a control parameter $x$. 
For centrosymmetric systems with $\mathcal{T}$-symmetry, there is generally an intermediate mixed-parity superconducting phase ($\Gamma_g+i\Gamma_u$) where even- and odd-parity pairing components are coexistent with the relative phase difference $\pm\pi/2$.
(b) Schematic figure of the anapole superconducting states. In real space, the phase of the superconducting order parameter becomes nonuniform along a direction parallel to the effective anapole moment $\bm{T}$ as $\Delta(\bm{r})\propto e^{i\bm{q}\cdot\bm{r}}$ with $\bm{q}\parallel\bm{T}$. (c) Illustration for a mechanism of the asymmetric BS in $\mathcal{PT}$-symmetric mixed-parity multiband superconductors. $\mathcal{P}$- and $\mathcal{T}$-symmetry breaking interband pairing induces an asymmetric modulation of the BS.}
\label{fig:phase}
\end{figure*}
In particular, {\it mixed-parity superconductivity} with coexistent even- and odd-parity pairing channels has been widely discussed in noncentrosymmetric superconductors~\cite{Bauer2012NCS,Smidman2017},
ultracold fermion systems~\cite{ultracold2010,Zhou2017}, and spin-orbit-coupled systems in the vicinity of the $\mathcal{P}$-symmetry breaking~\cite{Fu2015_SOC,Kozii-Fu2015,Sumita2020,Hiroi2018pyrochlore,STO_parity}. 
The $\mathcal{P}$-symmetry is broken in such superconductors, and spontaneous $\mathcal{T}$-symmetry breaking realized by the $\pm\pi/2$ phase difference between even- and odd-parity pairing potentials is energetically favored~\cite{WangFu2017,Wang2020} (Fig.~\ref{fig:phase}a) when the spin-orbit coupling (SOC) due to noncentrosymmetric crystal structure is absent or weak.
This class of superconducting states 
spontaneously breaks both $\mathcal{P}$- and $\mathcal{T}$-symmetries but maintain the combined $\mathcal{PT}$-symmetry. 
There have been considerable interests in studying such {\it $\mathcal{PT}$-symmetric mixed-parity superconductivity}. 
The three-dimensional $s+ip$-wave superconductivity has attracted much theoretical attention as a superconducting analog of axion insulators~\cite{Ryu2012,Qi2013,Goswami2014,Shiozaki2014,Roy2020,Xu2020axion}. 
The $\mathcal{T}$-symmetry breaking mixed-parity pairing has also been theoretically proposed in Sr$_2$RuO$_4$~\cite{SRO_mixed-parity}. 
Furthermore, a mixed-parity superconducting state in UTe$_2$~\cite{Ishizuka2020} has been predicted  to explain experimentally-observed multiple superconducting phases~\cite{Braithwaite2019,Lin2020,Aoki2020,aoki2021fieldinduced}.

In previous works, the mixed-parity superconductivity has been theoretically studied mainly in single-band models for spin-$1/2$ fermions~\cite{WangFu2017,Wang2020,Ryu2012,Qi2013,Goswami2014,Shiozaki2014,Xu2020axion}.  
On the other hand, it has recently been recognized that the multiband structure of the Cooper pair's wave function arising from internal electronic degrees of freedom (DOF) (e.g., orbital and sublattice) induces exotic superconducting phenomena. 
For instance, multiband superconductors have attracted much attention as a platform realizing odd-frequency pairing~\cite{AMB_AVB_2013}. 
In $\mathcal{T}$-symmetry breaking superconductors, an intrinsic anomalous Hall effect emerges owing to the multiband nature of Cooper pairs~\cite{Taylor2012AHE,Wang2017,Brydon2019AHE,Brydon2021AHE,Triola2018}. 
In particular, even-parity $\mathcal{T}$-symmetry breaking superconductors host topologically protected Bogoliubov Fermi surfaces in the presence of interband pairing~\cite{Agterberg_BogoFS,Brydon_BogoFS}. 

In this work, we show that $\mathcal{PT}$-symmetric mixed-parity superconducting states generally exhibit an asymmetric Bogoliubov spectrum (BS) in multiband systems, although it is overlooked in single-band models. 
We demonstrate that such asymmetric deformation of the BS is induced by a nonunitary interband pairing (see Fig.~\ref{fig:phase}c), and derive the necessary conditions 
for generic two-band models. 
Although we consider two-band systems for simplicity throughout this paper, our theory is relevant for any multiband superconductors with multiple bands near the Fermi level.  
In addition, we show that the Bogoliubov quasiparticles with asymmetric BS stabilize the Fulde-Ferrell-Larkin-Ovchinnikov (FFLO) superconductivity~\cite{FF1964,LO1964}, which is evident from the Lifshitz invariants~\cite{Mineev_Lifshitz}, namely linear gradient terms, in the Ginzburg-Landau (GL) free energy. 
The Lifshitz invariants are nonzero only for the {\it anapole superconducting states}, whose order parameters are equivalent to an anapole (magnetic toroidal) moment, namely a polar and time-reversal odd multipole~\cite{Spaldin_2008}, from the viewpoint of symmetry.
It is shown that the phase of the superconducting order parameter is spatially modulated along the effective anapole moment of the superconducting state (see Fig.~\ref{fig:phase}b). 
The concept of anapole order has been employed in nuclear physics~\cite{flambaum1984nuclear}, magnetic materials science~\cite{Spaldin_2008}, strongly correlated electron systems~\cite{Jeong2017,Murayama2021}, and optoelectronics~\cite{HW2021_Photocurrent,Ahn2020}, and it is extended to superconductors by this work. 
In previous works, the FFLO superconductivity has been proposed in the presence of an external magnetic field~\cite{FF1964,LO1964,Agterberg2007FFLO,Mineev_Lifshitz} or coexistent magnetic multipole order~\cite{Sumita2016,Sumita2017}. 
However, the magnetic field causes superconducting vortices, prohibiting pure FFLO states, and the proposed multipole superconducting state has not been established in condensed matters.
In contrast, the anapole superconductivity realizes the FFLO state without the aid of any other perturbation or electronic order. 
Note that an intrinsic nonuniform superconducting state has also been discussed in the Bogoliubov Fermi surface states~\cite{Timm2021Distortional}, although its mechanism and symmetry are different from those of the anapole FFLO state.

Based on the obtained results, we predict the possible asymmetric BS and anapole superconductivity in UTe$_2$, a recently-discovered candidate of a spin-triplet superconductor~\cite{Ran2019}. The multiple pairing instabilities~\cite{Braithwaite2019,Lin2020,Aoki2020,aoki2021fieldinduced}
and $\mathcal{T}$-symmetry breaking 
(e.g, Ref.~\cite{Jiao2020chiral})
were recently observed there.

\section*{Results}
\textbf{General two-band Bogoliubov-de Gennes Hamiltonian.}
We begin our discussion by considering the general form of the Bogoliubov-de Gennes (BdG) Hamiltonian for two-band systems:
\begin{align}
    \mathcal{H} = \frac{1}{2}\sum_{\bm{k}}
      (\hat{c}_{\bm{k}}^{\dag} , \hat{c}_{-\bm{k}}^{\rm T} )
    \begin{pmatrix}
      H_0(\bm{k}) & \Delta(\bm{k}) \\
       \Delta^{\dag}(\bm{k}) & -H_0^{*}(-\bm{k})
    \end{pmatrix}
    \begin{pmatrix}
      \hat{c}_{\bm{k}} \\
      \hat{c}_{-\bm{k}}^{*}
    \end{pmatrix} , 
\end{align}
where $\hat{c}_{\bm{k}}^{\rm T}=(c_{\bm{k}1\uparrow},c_{\bm{k}1\downarrow},c_{\bm{k}2\uparrow},c_{\bm{k}2\downarrow})$ is a spinor encoding the four internal electronic DOF stem from spin-$1/2$ and extra two-valued DOF, such as orbitals and sublattices. 
Then, the $4\times4$ matrices $H_0(\bm{k})$ and $\Delta(\bm{k})$ can be generally expressed as a linear combination of $\sigma_{\mu}\otimes\tau_{\nu}$ matrices, where $\sigma_{\mu}$ and $\tau_{\nu}$ ($\mu,\nu=0,x,y,z$) are the Pauli matrices for the spin and extra DOF, respectively. 
However, we here introduce a more convenient form of the two-band BdG Hamiltonian using the Euclidean Dirac matrices $\gamma_{n}$ ($n=1,2,3,4,5$), which satisfy $\{\gamma_m,\gamma_n\}=2\delta_{mn}$. See Methods section for the correspondence between the $\sigma_{\mu}\otimes\tau_{\nu}$ and Dirac matrices. 
Assuming that the normal state preserves both $\mathcal{P}$- and $\mathcal{T}$-symmetries, the general form of the normal state Hamiltonian $H_0(\bm{k})$ can be expressed as  
\begin{equation}
    H_{0}(\bm{k}) = (\epsilon^{0}_{\bm{k}}-\mu)\mathbbm{1}_{4} + \bm{\epsilon}_{\bm{k}}\cdot\bm{\gamma} ,
    \label{eq:H0-gamma}
\end{equation}
where $\mathbbm{1}_4$ is the $4\times4$ unit matrix, $\bm{\gamma}=(\gamma_1,\gamma_2,\gamma_3,\gamma_4,\gamma_5)$ is the vector of the five Dirac matrices, 
$\epsilon_{\bm{k}}^0$ and $\bm{\epsilon}_{\bm{k}}=(\epsilon_{\bm{k}}^{1},\epsilon_{\bm{k}}^{2},\epsilon_{\bm{k}}^{3},\epsilon_{\bm{k}}^{4},\epsilon_{\bm{k}}^{5})$ are the real-valued coefficients of these matrices, and $\mu$ is the chemical potential. 
Whereas $\epsilon_{\bm{k}}^0$ is an even function of momentum, $\bm{k}$-parity of other coefficients $\epsilon_{\bm{k}}^{n}$ ($n>0$) depends on the details of the extra DOF. 
The superconducting state is assumed to be a mixture of even- and odd-parity pairing components. 
The pairing potential $\Delta(\bm{k})$ for such mixed-parity superconducting states has the general form
\begin{equation}
    \hat{\Delta}(\bm{k}) = \Delta_1(\eta^{0}_{\bm{k}}\mathbbm{1}_{4} + \bm{\eta}_{\bm{k}}\cdot\bm{\gamma})
    + \Delta_2\sum_{m<n}\eta^{mn}_{\bm{k}}i\gamma_{m}\gamma_{n} ,
    \label{eq:Delta-gamma}
\end{equation}
where $\hat{\Delta}(\bm{k})\equiv\Delta(\bm{k})U_T^{\dag}$ and $U_{T}$ is the unitary part of the time-reversal operator. 
The complex-valued constants $\Delta_1$ and $\Delta_2$ represent the superconducting order parameters for the even- and odd-parity pairing channels, respectively. 
As a consequence of the fermionic antisymmetry $\Delta(\bm{k})=-\Delta^{\rm T}(-\bm{k})$, the even-parity (odd-parity) part of $\hat{\Delta}(\bm{k})$ is expressed by a linear combination of $\mathbbm{1}_4$ and $\gamma_n$ ($i\gamma_m\gamma_n$) as shown in Equation~\eqref{eq:Delta-gamma} (see Methods). 
The real-valued functions $\eta_{\bm{k}}^0$, $\bm{\eta}_{\bm{k}}=(\eta_{\bm{k}}^{1}, \eta_{\bm{k}}^{2},\eta_{\bm{k}}^{3},\eta_{\bm{k}}^{4},\eta_{\bm{k}}^{5})$, and $\eta_{\bm{k}}^{mn}$ ($1\leq m<n \leq5$) 
determines the details of order parameters.
Whereas $\eta_{\bm{k}}^0$ is an even function of momentum, $\bm{k}$-parity of others $\eta_{\bm{k}}^{n}$ and $\eta_{\bm{k}}^{mn}$ depends on the details of the extra DOF.
Note that the $\bm{k}$-parity of $\eta_{\bm{k}}^{n}$ must be the same as that of $\epsilon_{\bm{k}}^{n}$.  
\\

\textbf{Asymmetric BS from $\mathcal{PT}$-symmetric mixed-parity interband pairing.}
In the following, we assume that the intraband pairing is dominant compared to the interband pairing. 
In such situations, spontaneous $\mathcal{T}$-symmetry breaking with maintaining the $\mathcal{PT}$-symmetry is energetically favored in the mixed-parity superconducting states~\cite{WangFu2017,Wang2020}, and the symmetry of the superconducting order parameter becomes equivalent to that of odd-parity magnetic multipoles~\cite{Spaldin_2008}.  
A characteristic feature of the odd-parity magnetic multipole ordered state is the asymmetric modulation of the band structure~\cite{Yanase2014,Hayami2014,Sumita2016,Sumita2017}, which leads to peculiar nonequilibrium responses such as nonreciprocal transport~\cite{Rikken2001-il}, magnetopiezoelectric effect~\cite{HW2017_magnetopiezo,Shiomi2019}, and photocurrent generation~\cite{Ahn2020,HW2021_Photocurrent}.
Therefore, the appearance of the asymmetric BS is naturally expected in the $\mathcal{PT}$-symmetric mixed-parity superconductors. 
However, the asymmetric BS is not obtained in single-band models (see later discussions).

To induce such asymmetric modulation in the BS, effects of the $\mathcal{P}$- and $\mathcal{T}$-symmetry breaking in the particle-particle superconducting channel should be transferred into the particle-hole channel. 
This suggests that it is not sufficient to consider only the pairing potential $\Delta(\bm{k})$, since it is not gauge invariant. 
Instead of $\Delta(\bm{k})$ alone, we need to consider gauge-invariant bilinear products of $\Delta(\bm{k})$ and $\Delta^{\dag}(\bm{k})$~\cite{Brydon2019AHE} in order to reveal conditions for realizing the asymmetric BS. 
Here, we focus on the simplest bilinear products, that is, $\Delta(\bm{k})\Delta^{\dag}(\bm{k})$. 
The parity-odd and time-reversal-odd ($\mathcal{P,T}$-odd) part of this bilinear product 
is calculated as 
\begin{equation}
    M_{-}^{(1)}(\bm{k}) = \frac{1}{2}\left([\hat{\Delta}^{g}(\bm{k}),\hat{\Delta}^{u\dag}(\bm{k})]+[\hat{\Delta}^{u}(\bm{k}),\hat{\Delta}^{g\dag}(\bm{k})]\right),
    \label{eq:parity-time-odd-gap-prod}
\end{equation}
where $\hat{\Delta}^{g}(\bm{k})$ and $\hat{\Delta}^{u}(\bm{k})$ are the even- and odd-parity part of $\hat{\Delta}(\bm{k})$, respectively (see Supplementary Information).
Owing to the gauge invariance and $\mathcal{P},\mathcal{T}$-odd behavior of $M_{-}^{(1)}(\bm{k})$, a nonzero $M_{-}^{(1)}(\bm{k})$ can be a measure of the $\mathcal{P}$- and $\mathcal{T}$-symmetry breaking in the particle-hole channel, which permits emergence of the asymmetric BS. 
Note that the pairing state must be nonunitary to induce a nonzero $M_{-}^{(1)}(\bm{k})$, since $M_{-}^{(1)}(\bm{k})=0$ when $\Delta(\bm{k})\Delta^{\dag}(\bm{k})$ is proportional to the unit matrix. 
In analogy with the spin polarization of nonunitary spin-triplet superconducting states in spin-1/2 single-band models~\cite{Sigrist-Ueda}, the $\mathcal{P,T}$-odd bilinear product $M_{-}^{(1)}(\bm{k})$ can be interpreted as a polarization of an internal DOF in the superconducting state.

The emergence of a nonzero $\mathcal{P,T}$-odd bilinear product $M_{-}^{(1)}(\bm{k})$ requires the interband pairing. 
To see this, we consider the problem in the band basis.
Since $H_0(\bm{k})$ is assumed to preserve the $\mathcal{P}$- and $\mathcal{T}$-symmetries, the energy eigenvalues are doubly degenerate and labelled by a pseudospin index. Especially, we choose the so-called manifestly covariant Bloch basis~\cite{Fu2015_SOC}, in which the pseudospin index transforms like a true spin-1/2 under time-reversal and crystalline symmetry operations. 
In this basis, the intraband pairing potential is generally expressed as $\Delta_{\bm{k}} = (\psi_{\bm{k}}+\bm{d}_{\bm{k}}\cdot\bm{s})is_y$, where $\bm{s}=(s_x,s_y,s_z)$ are Pauli matrices in pseudospin space.
The complex-valued functions $\psi_{\bm{k}}$ and $\bm{d}_{\bm{k}}$ are even and odd functions of $\bm{k}$, respectively. 
Then, in the absence of the interband pairing, the multiband BdG Hamiltonian matrix reduces to a series of decoupled blocks describing spin-$1/2$ single-band superconductors. 
The bilinear product for this intraband pairing potential is obtained as $\Delta_{\bm{k}}\Delta_{\bm{k}}^{\dag} = (|\psi_{\bm{k}}|^2+|\bm{d}_{\bm{k}}|^2) \mathbbm{1}_{2} + 2\mathrm{Re}(\psi_{\bm{k}}\bm{d}_{\bm{k}}^{*})\cdot\bm{s} + i(\bm{d}_{\bm{k}}\times\bm{d}_{\bm{k}}^{*})\cdot\bm{s}$,  
and the second and third terms are nonunitary components that break $\mathcal{P}$- and $\mathcal{T}$-symmetries, respectively. 
However, there appears no term breaking both $\mathcal{P}$- and $\mathcal{T}$-symmetries, and hence the $\mathcal{P,T}$-odd bilinear product for this $\Delta_{\bm{k}}$ must vanish. 
This indicates that the interband pairing is necessary for a nonzero $\mathcal{P,T}$-odd bilinear product $M_{-}^{(1)}(\bm{k})$, which is essential for realizing the asymmetric BS. 
This is also the reason the asymmetric BS is not obtained in single-band models. 

The presence of interband pairing can be characterized by the so-called superconducting fitness $F(\bm{k})$, which is defined as $F(\bm{k})U_T = H_0(\bm{k})\Delta(\bm{k})-\Delta(\bm{k})H_0^{*}(-\bm{k})$~\cite{SCF2016,SCF2018}. 
Since a nonvanishing $F(\bm{k})F^{\dag}(\bm{k})$ quantifies the strength of interband pairing by definition~\cite{SCF2016,SCF2018}, its $\mathcal{P,T}$-odd part should be nonzero to realize a nonvanishing $M_{-}^{(1)}(\bm{k})$. 
The $\mathcal{P,T}$-odd part of $F(\bm{k})F^{\dag}(\bm{k})$ is obtained as 
\begin{equation}
    M_{-}^{(2)}(\bm{k}) = \frac{1}{2}\left([F^{g}(\bm{k}),F^{u\dag}(\bm{k})]+[F^{u}(\bm{k}),F^{g\dag}(\bm{k})]\right),
    \label{eq:SCF-PTodd}
\end{equation}
where $F^{g}(\bm{k})$ and $F^{u}(\bm{k})$ are the even- and odd-parity part of $F(\bm{k})$, respectively. 
Note that the $\mathcal{P,T}$-odd part of $F(\bm{k})F^{\dag}(\bm{k})$ can be extracted in the same way as $\Delta(\bm{k})\Delta^{\dag}(\bm{k})$ [compare Equation~\eqref{eq:SCF-PTodd} with Equation~\eqref{eq:parity-time-odd-gap-prod}], since the transformation properties of $F(\bm{k})F^{\dag}(\bm{k})$ under space-inversion and time-reversal are the same as $\Delta(\bm{k})\Delta^{\dag}(\bm{k})$. 
Based on Equation~\eqref{eq:SCF-PTodd}, not only the pair potential $\Delta(\bm{k})$ but also the normal part $H_0(\bm{k})$ must satisfy a proper condition to realize $M_{-}^{(2)}(\bm{k})\neq0$ and asymmetric BS.

From the above discussions, we conclude that the necessary (but not sufficient) condition for the asymmetric BS can be written as $M_{-}^{(1)}(\bm{k})\neq0 \,\cap\, M_{-}^{(2)}(\bm{k})\neq0$, which implies the superconductivity-driven $\mathcal{P}$- and $\mathcal{T}$-symmetry breaking in the particle-hole channel.
We here write down this necessary condition for the general two-band BdG Hamiltonian.
By substituting Equations~\eqref{eq:H0-gamma} and \eqref{eq:Delta-gamma} to Equations~\eqref{eq:parity-time-odd-gap-prod} and \eqref{eq:SCF-PTodd}, we obtain the $\mathcal{P,T}$-odd bilinear products $M_{-}^{(1)}(\bm{k})$ and $M_{-}^{(2)}(\bm{k})$ 
as follows: 
\begin{align}
    M_{-}^{(1)}(\bm{k}) &=
    2\mathrm{Im}(\Delta_{1}\Delta_{2}^*)\sum_{m<n}\eta_{\bm{k}}^{mn}(\eta_{\bm{k}}^{n}\gamma_{m}-\eta_{\bm{k}}^{m}\gamma_{n}),
    \label{eq:PT-odd-gap-prod-2band}
    \\
    M_{-}^{(2)}(\bm{k}) &= \mathrm{Tr}[M_{-}^{(1)}(\bm{k})\tilde{H}_{0}(\bm{k})]\tilde{H}_{0}(\bm{k}),
    \label{eq:SCF-PTodd-2band}
\end{align}
where $\tilde{H}_0(\bm{k})\equiv H_0(\bm{k})-(\epsilon_{\bm{k}}^{0}-\mu)\mathbbm{1}_{4}$. 
Note that Equation~\eqref{eq:SCF-PTodd-2band} is derived for the general two-band model, but 
it may not hold in multiband models with more than two bands.
From Equations~\eqref{eq:PT-odd-gap-prod-2band} and \eqref{eq:SCF-PTodd-2band}, the necessary conditions for the asymmetric BS in two-band models can be summarized as following two criteria; 
(i) the relative phase difference between even- and odd-parity pairing potentials must be nonzero so that 
$\mathrm{Im}(\Delta_1\Delta_2^{*})\neq0$, 
and (ii) 
the BdG Hamiltonian must satisfy $\epsilon_{\bm{k}}^{m}\eta_{\bm{k}}^{n}\eta_{\bm{k}}^{mn}\neq0$ or $\epsilon_{\bm{k}}^{n}\eta_{\bm{k}}^{m}\eta_{\bm{k}}^{mn}\neq0$ for $1\leq {}^\exists m< {}^\exists n\leq5$ (i.e., $\mathrm{Tr}[M_{-}^{(1)}(\bm{k})\tilde{H}_{0}(\bm{k})]\neq0$). 
Interpretations of these requirements in the $\sigma_{\mu}\otimes\tau_{\nu}$ basis are shown in Methods section. 

We now confirm that the asymmetric BS indeed appears when the above two criteria are fulfilled. 
A minimal two-band model satisfying the criterion (ii) can be obtained by substituting $\bm{\epsilon}_{\bm{k}}=r\epsilon_{\bm{k}}^{a}\bm{e}_{a}+(1-r)\epsilon_{\bm{k}}^{b}\bm{e}_{b}$, $\bm{\eta}_{\bm{k}}=(1-r)\eta_{\bm{k}}^{a}\bm{e}_{a}+r\eta_{\bm{k}}^{b}\bm{e}_{b}$, and $\eta_{\bm{k}}^{mn}=\delta_{ma}\delta_{nb}\eta_{\bm{k}}^{ab}$ into Equations~\eqref{eq:H0-gamma} and \eqref{eq:Delta-gamma}. 
Here, $a$ and $b$ are specific integers satisfying $1\leq a < b\leq 5$, $\bm{e}_{n}$ is the unit vector for the $n$-th component, and $r$ takes the value either $0$ or $1$. 
Under this setup, we can analytically diagonalize the BdG Hamiltonian as $\mathrm{diag}(E_{\bm{k}}^{+}\mathbbm{1}_2,E_{\bm{k}}^{-}\mathbbm{1}_2,-E_{-\bm{k}}^{+}\mathbbm{1}_2,-E_{-\bm{k}}^{-}\mathbbm{1}_2)$, where $\mathbbm{1}_2$ is the $2\times2$ unit matrix. 
Based on the correspondence between the Dirac matrices and $\sigma_{\mu}\otimes\tau_{\nu}$ matrices 
the energy spectrum $E_{\bm{k}}^{\pm}$ can be obtained as 
\begin{align}
 E_{\bm{k}}^{\pm} &= \sqrt{\xi_{\bm{k}}^2+\frac{1}{4}\mathrm{Tr}\left[\Delta(\bm{k})\Delta^{\dag}(\bm{k})\pm\frac{M_{-}^{(1)}(\bm{k})\tilde{H}_0(\bm{k})}{r\epsilon_{\bm{k}}^{a}+(1-r)\epsilon_{\bm{k}}^{b}}\right]} \nonumber\\
 & \pm [r\epsilon_{\bm{k}}^{a}+(1-r)\epsilon_{\bm{k}}^{b}],
 \label{eq:spectrum}
\end{align}
where $\xi_{\bm{k}}\equiv\epsilon_{\bm{k}}^0-\mu$.
By using the transformation properties of the BdG Hamiltonian under space-inversion and time-reversal, we can confirm that Equation~\eqref{eq:spectrum} satisfies $E_{-\bm{k}}^{+} \neq E_{\bm{k}}^{\pm}$ and $E_{-\bm{k}}^{-} \neq E_{\bm{k}}^{\pm}$ (i.e., the BS is asymmetric) when $\mathrm{Tr}[M_{-}^{(1)}(\bm{k})\tilde{H}_{0}(\bm{k})]\neq0$. See Methods for the proof. 
This implies that $M_{-}^{(1)}(\bm{k})\neq0 \,\cap\, M_{-}^{(2)}(\bm{k})\neq0$ is indeed a necessary condition for emergence of the asymmetric BS. \\

\textbf{Lifshitz invariants and effective anapole moment.}
To obtain further insight into the asymmetric BS, we now investigate the free energy of the above minimal model satisfying $M_{-}^{(1)}(\bm{k})\neq0 \,\cap\, M_{-}^{(2)}(\bm{k})\neq0$. 
By differentiating Equation~\eqref{eq:spectrum} with respect to $\Delta_j$ and $\Delta_j^*$ ($j=1,2$), the GL free energy for superconductivity is derived as follows (see Supplementary Information):
\begin{align}
 \mathcal{F} &= \alpha_{1}|\Delta_{1}|^2 + \alpha_{2}|\Delta_{2}|^2 
 + \beta_{1}|\Delta_{1}|^4 + \beta_{2}|\Delta_{2}|^4 \nonumber\\
 & + 4\tilde{\beta}|\Delta_{1}|^2|\Delta_{2}|^2  -\tilde{\beta}(\Delta_{1}^{2}\Delta_{2}^{*2}+\Delta_{2}^{2}\Delta_{1}^{*2})
 \nonumber\\
& + \sum_{\nu=x,y,z}(\kappa_{1,\nu}|\Delta_{1}|^{2}+\kappa_{2,\nu}|\Delta_{2}|^{2})q_{\nu}^{2} + \bm{T}\cdot\bm{q}, \label{eq:GL_free_energy}
\end{align}
where $\bm{q}=(q_x, q_y, q_z)$ is the center-of-mass momentum of Cooper pairs. 
The analytical expressions of $\alpha_{j}$, $\beta_{j}(>0)$, $\tilde{\beta}(>0)$, and $\kappa_{j,\nu}(>0)$ are shown in Supplemental Information.
The last term is the Lifshitz invariant~\cite{Mineev_Lifshitz} stabilizing the FFLO state with $\bm{q}\parallel\bm{T}$. 
Since the Cooper pair condensation occurs at a single $\bm{q}$ in our model, the superconducting order parameter is expressed as $\Delta(\bm{r})\propto e^{i\bm{q}\cdot\bm{r}}$ in real space (Fig.~\ref{fig:phase}b).
The coefficient vector $\bm{T}=(T_x,T_y,T_z)$ is given by
\begin{equation}
    \bm{T} = \rho_0\average{\mathrm{Tr}[M_{-}^{(1)}(\bm{k})\tilde{H}_{0}(\bm{k})]\bm{v}_{\bm{k}}}_{\rm FS}\frac{7\zeta(3)}{16\pi^2T^2},
    \label{eq:Lifshitz-invariant}
\end{equation}
where $\rho_0$ is the density of states at the Fermi energy, $\average{\cdots}_{\rm FS}$ denotes the average over the Fermi surface, $\bm{v}_{\bm{k}}\equiv\nabla_{\bm{k}}\xi_{\bm{k}}$, $T$ is the temperature, and $\zeta(x)$ is the Riemann zeta function.
$\bm{T}$ can be interpreted as the effective anapole moment of the superconducting state. 
To see this, we here consider conditions for $\bm{T}\neq\bm{0}$. 
Equation~\eqref{eq:Lifshitz-invariant} indicates that $\bm{T}$ is nonzero only for $\mathcal{P}$- and $\mathcal{T}$-symmetry breaking pairing states with $M_{-}^{(1)}(\bm{k})\neq0$. 
In addition, $\average{\mathrm{Tr}[M_{-}^{(1)}(\bm{k})\tilde{H}_{0}(\bm{k})]\bm{v}_{\bm{k}}}_{\rm FS}$ is nonzero only when the superconducting order parameter belongs to a polar irreducible representation (IR), since the velocity $\bm{v}_{\bm{k}}$ is a polar vector and $\tilde{H}_{0}(\bm{k})$ is assumed to be $\mathcal{P}$-symmetric. 
Therefore, $\bm{T}$ is a polar and time-reversal-odd vector; the symmetry is equivalent to the anapole moment~\cite{Spaldin_2008,flambaum1984nuclear}. 
Hereafter, we refer to the superconductivity with $\bm{T}\neq\bm{0}$ as the {\it anapole superconductivity}.
The anapole superconductivity realizes a nonuniform FFLO state with $\bm{q}\parallel\bm{T}$ (see Fig.~\ref{fig:phase}b) to compensate a polar asymmetry in the BS. \\  

\textbf{Application to UTe$_2$.}
We now discuss the asymmetric BS and anapole superconductivity in UTe$_2$. Intensive studies after the discovery of superconductivity evidenced odd-parity spin-triplet superconductivity in UTe$_2$~\cite{Ran2019,Aoki2019,Knafo2019RSC,Metz2019thermal,Nakamine2021}. However, multiple superconducting phases similar to Fig.~\ref{fig:phase}a have been observed under pressure~\cite{Braithwaite2019,Lin2020,Aoki2020,aoki2021fieldinduced}, and the antiferromagnetic quantum criticality implies the spin-singlet superconductivity there~\cite{Thomas2020}. A theoretical study based on the periodic Anderson model verified this naive expectation and predicted the parity-mixed superconducting state in the intermediate pressure region~\cite{Ishizuka2020}. 

\begin{figure}[tbp]
\begin{centering}
\includegraphics[width=1\linewidth]{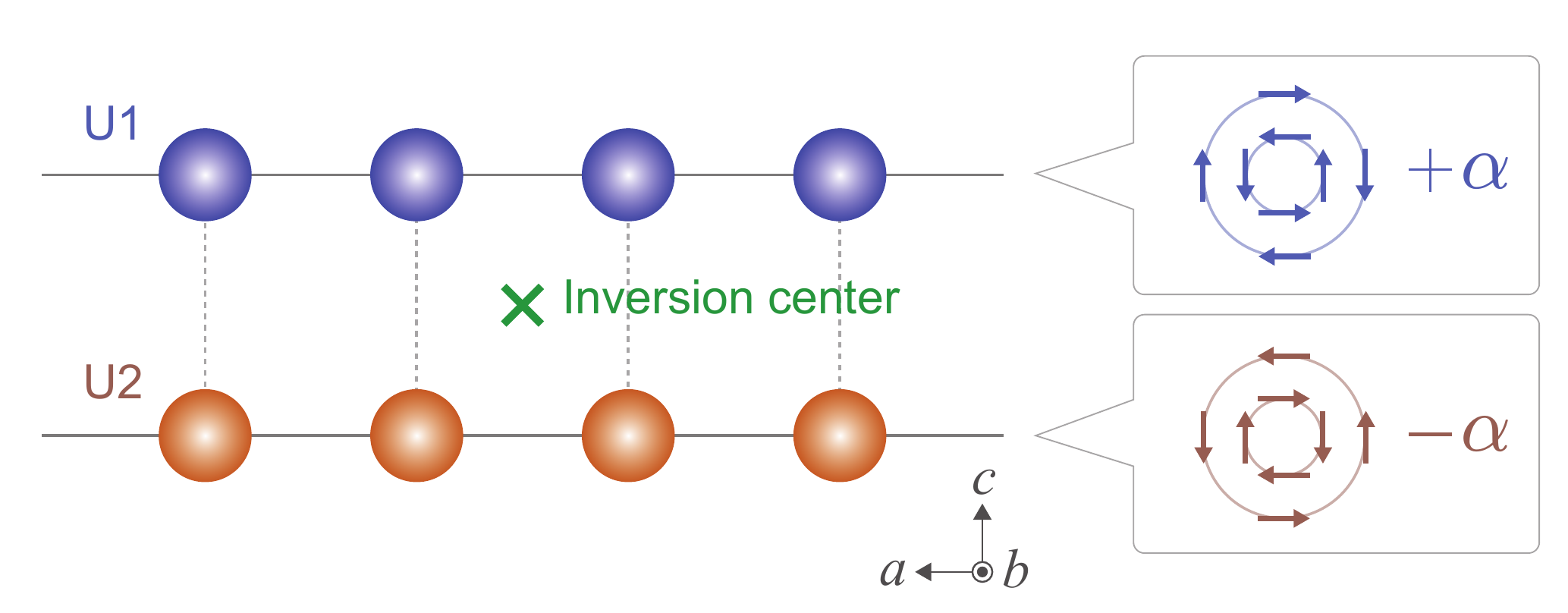}
\end{centering}	
\caption{\textbf{Schematic of local $\mathcal{P}$-symmetry breaking in UTe$_2$.} 
U atoms form a ladder structure along the $a$-axis. This locally noncentrosymmetric crystal structure leads to a sublattice-dependent staggered form of Rashba-type SOC. }
\label{fig:UTe2}
\end{figure}
Since the crystal structure of UTe$_2$ preserves $D_{2h}$ point group symmetry, the superconducting order parameter is classified based on the IRs of $D_{2h}$. 
We consider all the odd-parity $A_{u}$, $B_{1u}$, $B_{2u}$, and $B_{3u}$ IRs, although the $A_{u}$ and $B_{3u}$ IRs may be promising candidates 
based on the experimental results and theoretical arguments~\cite{Ishizuka2020,Ishizuka2019,Metz2019thermal,Nakamine2021,Hiranuma2021}.
On the other hand, a recent calculation has shown that the even-parity $A_{g}$ superconducting state is favored by antiferromagnetic fluctuation under pressure~\cite{Ishizuka2020}. 
Therefore, the superconductivity at an intermediate pressure regime is expected to be a mixture of the even-parity $A_{g}$ and odd-parity $A_u$ or $B_{3u}$ states.
Note that the interband pairing, which is an essential ingredient for the asymmetric BS and anapole superconductivity, may have considerable impacts on the superconductivity in UTe$_2$ owing to multiple bands near the Fermi level (see e.g., Refs.~\cite{Ishizuka2019,Xu2019DMFT}).

Based on the above facts, we introduce a minimal model for UTe$_2$ as follows: 
\begin{align}
    H_0(\bm{k}) &= (\varepsilon_{\bm{k}}-\mu)\sigma_0\otimes\tau_0+ \bm{g}_{\bm{k}}\cdot\bm{\sigma}\otimes\tau_z,
    \label{eq:minimal_H0}
    \\
    \hat{\Delta}(\bm{k}) &= \Delta_{1}(\psi_{\bm{k}}^{g}\sigma_0\otimes\tau_0+\bm{d}^{g}_{\bm{k}}\cdot\bm{\sigma}\otimes\tau_z) \nonumber\\
    & + \Delta_{2}(\bm{d}^{u}_{\bm{k}}\cdot\bm{\sigma}\otimes\tau_0 + \psi_{\bm{k}}^{u}\sigma_0\otimes\tau_z), 
    \label{eq:minimal_Delta}
\end{align}
where $\tau_{\nu}$ represents the Pauli matrix for a sublattice DOF originating from a ladder structure of U atoms (Fig.~\ref{fig:UTe2}). 
We assume a simple form of the single-particle kinetic energy  as $\varepsilon_{\bm{k}}=-2\sum_{\nu=x,y,z}t_{\nu}\cos k_{\nu}$.
The second term of Equation (\ref{eq:minimal_H0}) is a sublattice-dependent staggered form of Rashba SOC with $\bm{g}_{\bm{k}}=\alpha(\sin{k_y}\hat{\bm{x}}-\sin{k_x}\hat{\bm{y}})$, which originates from the local $\mathcal{P}$-symmetry breaking at U sites~\cite{Ishizuka2020,Shishidou2021}. 
Since the local site symmetry descends to $C_{2v}$ from $D_{2h}$ owing to the ladder structure of U atoms, the existence of the Rashba-type SOC with opposite coupling constants $\pm\alpha$ at each sublattices is naturally expected (see Fig.~\ref{fig:UTe2}).
The local $\mathcal{P}$-symmetry breaking also leads to a sublattice-dependent parity mixing of the pair potential~\cite{Fischer2011}. 
Then, the even-parity (odd-parity) pair potential is assumed to be a mixture of intrasublattice spin-singlet (spin-triplet) and staggered spin-triplet (spin-singlet) components as shown in Equation~\eqref{eq:minimal_Delta}. 
We assume the form of the $\bm{k}$-dependent coefficients $\psi_{\bm{k}}^{g}$ and $\bm{d}_{\bm{k}}^{g}$ ($\bm{d}_{\bm{k}}^{u}$ and $\psi_{\bm{k}}^{u}$) so as to be consistent with the basis functions of the $A_g$ IR ($A_u$, $B_{1u}$, $B_{2u}$, or $B_{3u}$ IRs). 

\begin{table*}[htbp] 
\caption{\label{tab:UTe2}
Basis functions of $\bm{d}_{\bm{k}}^{g,u}$ and corresponding $\bm{g}_{\bm{k}}\cdot(\bm{d}^{g}_{\bm{k}}\times\bm{d}_{\bm{k}}^{u})$ for possible $\mathcal{PT}$-symmetric mixed-parity pairing states in UTe$_2$. The last column shows the form of the effective anapole moment $\bm{T}$ for each pairing state. 
}
\centering
\begin{ruledtabular}
{\renewcommand \arraystretch{1.3}
 \begin{tabular}{ccccc} 
Pairing & $\bm{d}_{\bm{k}}^{g}$ & $\bm{d}_{\bm{k}}^{u}$ & $\bm{g}_{\bm{k}}\cdot(\bm{d}^{g}_{\bm{k}}\times\bm{d}_{\bm{k}}^{u})$ & $\bm{T}$
\\ \hline
$A_g+iA_u$ & $\phi_{x}^g k_y\hat{\bm{x}}+\phi_{y}^gk_x\hat{\bm{y}}$ & $\phi_x^uk_x\hat{\bm{x}}+\phi_y^uk_y\hat{\bm{y}}+\phi_z^uk_z\hat{\bm{z}}$ & $\alpha(\phi_x^g+\phi_y^g)\phi_z^u k_xk_yk_z$ & $\bm{T}=\bm{0}$ \\
$A_g+iB_{3u}$ & $\phi_{x}^gk_y\hat{\bm{x}}+\phi_{y}^gk_x\hat{\bm{y}}$ & $\phi_y^uk_z\hat{\bm{y}}+\phi_z^uk_y\hat{\bm{z}}$ & $\alpha(\phi_x^g+\phi_y^g)\phi_z^u k_xk_y^2$ & $\bm{T}\parallel\hat{\bm{x}}$ \\
$A_g+iB_{2u}$ & $\phi_{x}^gk_y\hat{\bm{x}}+\phi_{y}^gk_x\hat{\bm{y}}$ & $\phi_x^uk_z\hat{\bm{x}}+\phi_z^uk_x\hat{\bm{z}}$ & $\alpha(\phi_x^g+\phi_y^g)\phi_z^u k_x^2k_y$ & $\bm{T}\parallel\hat{\bm{y}}$ \\
$A_g+iB_{1u}$ & $\phi_{x}^gk_y\hat{\bm{x}}+\phi_{y}^gk_x\hat{\bm{y}}$ & $\phi_x^uk_y\hat{\bm{x}}+\phi_y^uk_x\hat{\bm{y}}+\phi_z^uk_xk_yk_z\hat{\bm{z}}$ & $\alpha(\phi_x^g+\phi_y^g)\phi_z^u k_x^2k_y^2k_z$ & $\bm{T}\parallel\hat{\bm{z}}$
\end{tabular}
}
\end{ruledtabular}
\end{table*}
We now consider the necessary conditions for an asymmetric BS in UTe$_2$. 
As discussed in the above sections, a nonzero $\mathrm{Tr}[M_{-}^{(1)}(\bm{k})\tilde{H}_0(\bm{k})]$ is necessary for the asymmetric BS in a two-band model. 
For Equations~\eqref{eq:minimal_H0} and \eqref{eq:minimal_Delta}, this quantity is obtained as $\mathrm{Tr}[M_{-}^{(1)}(\bm{k})\tilde{H}_0(\bm{k})]=-8\mathrm{Im}(\Delta_1\Delta_2^*)[\bm{g}_{\bm{k}}\cdot(\bm{d}_{\bm{k}}^{g}\times\bm{d}_{\bm{k}}^{u})]$. 
Therefore, $\bm{g}_{\bm{k}}\cdot(\bm{d}_{\bm{k}}^{g}\times\bm{d}_{\bm{k}}^{u})\neq0$ must be satisfied to realize the asymmetric BS. 
This indicates that the sublattice-dependent SOC and spin-triplet pairing components $\bm{d}_{\bm{k}}^{g,u}$ are essential for the appearance of the asymmetric BS.  
On the other hand, the spin-singlet pairing components $\psi_{\bm{k}}^{g,u}$ do not play an important role for realizing the asymmetric BS in this model. 
Hereafter, we assume $\psi_{\bm{k}}^g=1$ and $\psi_{\bm{k}}^u=0$ for simplicity. 
The basis functions of $\bm{d}_{\bm{k}}^{g,u}$ and corresponding $\bm{g}_{\bm{k}}\cdot(\bm{d}_{\bm{k}}^{g}\times\bm{d}_{\bm{k}}^{u})$ for possible mixed-parity superconducting states in UTe$_2$ are summarized in Table \ref{tab:UTe2}. 
As shown in Table \ref{tab:UTe2}, $\bm{g}_{\bm{k}}\cdot(\bm{d}_{\bm{k}}^{g}\times\bm{d}_{\bm{k}}^{u})\propto\alpha(\phi_x^g+\phi_y^g)\phi_z^u$ for all patterns of the superconducting state, where $\phi_{\nu}^{g,u}$ is a real-valued coefficient of the $\nu$-th component of $\bm{d}_{\bm{k}}^{g,u}$. 
Therefore, $\phi_x^g+\phi_y^g\neq0$ and $\phi_z^u\neq0$ are necessary for the asymmetric BS. 
According to a recent numerical calculation in Ref.~\cite{Ishizuka2020}, the magnetic anisotropy of UTe$_2$ leads to $|\phi_y^{g}|\gg|\phi_x^{g}|$ for the $A_{g}$ state. 
Then, we assume $\bm{d}_{\bm{k}}^{g}=\sin{k_x}\hat{\bm{y}}$ (i.e., $\phi_x^g=0$ and $\phi_y^g=1$) in the following calculations. 
On the other hand, we assume $\phi_\nu^u=\delta_{\nu z}$ for the odd-parity pairing component to extract only the essential ingredient for the asymmetric BS and make a clear discussion. 

\begin{figure*}[tbp]
\begin{centering}
\includegraphics[width=\linewidth]{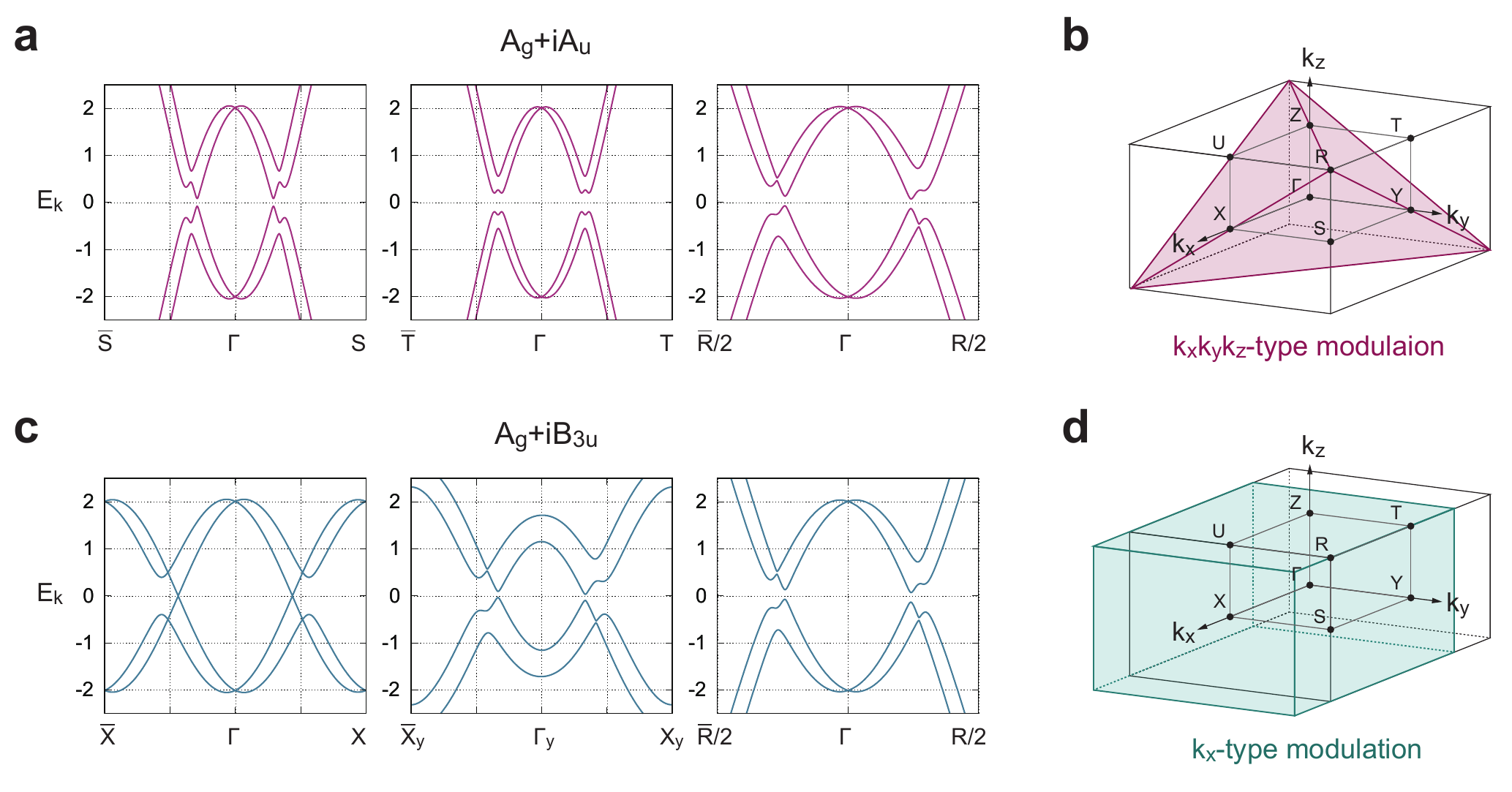}
\end{centering}	
\caption{\textbf{Asymmetric BS for the UTe$_2$ model.} 
(a) BS in the $A_{g}+iA_{u}$ state with $(\Delta_{1}, \Delta_{2})=(0.2, 0.2i)$ and $\bm{d}_{\bm{k}}^{u}=\sin{k_z}\hat{\bm{z}}$. 
(b) Schematic of $k_xk_yk_z$-type asymmetric modulation in the Brillouin zone. 
(c) BS in the $A_{g}+iB_{3u}$ state with $(\Delta_{1}, \Delta_{2})=(0.2, 0.2i)$ and $\bm{d}_{\bm{k}}^{u}=\sin{k_y}\hat{\bm{z}}$. 
(d) Schematic of $k_x$-type asymmetric modulation. 
The symbols of the horizontal axis in (a) and (c) denote the $\bm{k}$-points in the Brillouin zone of a primitive orthorhombic lattice; $\Gamma=(0,0,0)$, $\mathrm{X}=-\bar{\mathrm{X}}=(\pi,0,0)$,  $\mathrm{S}=-\bar{\mathrm{S}}=(\pi,\pi,0)$, $\mathrm{T}=-\bar{\mathrm{T}}=(0,\pi,\pi)$, $\mathrm{R}=-\bar{\mathrm{R}}=(\pi,\pi,\pi)$, $\Gamma_y=(0,\pi/4,0)$, and $\mathrm{X}_y=-\bar{\mathrm{X}}_y=(\pi,\pi/4,0)$. 
In the numerical calculations, parameters are set to be $t_{\nu}=1.0$, $\mu=-4.0$, and $\alpha=0.4$. 
The asymmetric BS appears in (a) and (c) consistent with the symmetry analysis of $\mathrm{Tr}[M_{-}^{(1)}(\bm{k})\tilde{H}_0(\bm{k})]\propto\bm{g}_{\bm{k}}\cdot(\bm{d}_{\bm{k}}^{g}\times\bm{d}_{\bm{k}}^{u})$. }
\label{fig:spectrum}
\end{figure*}
The numerical results of the BS for this UTe$_2$ model are shown in Fig.~\ref{fig:spectrum}. 
We here consider only the $A_{g}+iA_{u}$ and $A_{g}+iB_{3u}$ states as promising candidates of the $\mathcal{PT}$-symmetric mixed-parity superconductivity in UTe$_2$. 
It is shown that the BS of both $A_{g}+iA_{u}$ and $A_{g}+iB_{3u}$ states are indeed asymmetric along some directions in the Brillouin zone (see Figs.~\ref{fig:spectrum}a and \ref{fig:spectrum}c).
The BS in the $A_{g}+iA_{u}$ state exhibits a $k_xk_yk_z$-type tetrahedral asymmetry as depicted in Fig.~\ref{fig:spectrum}b, while the BS in the $A_{g}+iB_{3u}$ state shows a $k_xk_y^2$-type unidirectional asymmetry as depicted in Fig.~\ref{fig:spectrum}d.
Consistent with these numerical results, Table~\ref{tab:UTe2} reveals that $\bm{g}_{\bm{k}}\cdot(\bm{d}_{\bm{k}}^{g}\times\bm{d}_{\bm{k}}^{u})$ of the $A_{g}+iA_{u}$ and $A_{g}+iB_{3u}$ states are proportional to $k_xk_yk_z$ and $k_xk_y^2$, respectively.
This implies that the type of asymmetry in the BS is determined by the symmetry of $\mathrm{Tr}[M_{-}^{(1)}(\bm{k})\tilde{H}_0(\bm{k})]$, which is an essential ingredient for realizing the asymmetric BS. 

Finally, we discuss the possible anapole superconductivity in UTe$_2$. 
The $A_{g}+iA_{u}$ state belongs to the nonpolar $A_{u}^{-}$ IR (IRs with odd time-reversal parity are denoted by $\Gamma^{-}$), which corresponds to nonpolar odd-parity magnetic multipoles such as magnetic monopole, quadrupole, and hexadecapole from the viewpoint of symmetry. 
On the other hand, the $A_{g}+iB_{3u}$ state belongs to the polar $B_{3u}^{-}$ IR with the polar $x$-axis, which is symmetrically equivalent to the anapole moment $T_x$. 
Since the anapole superconducting states are allowed only when the superconducting order parameter belongs to a polar IR, the $A_{g}+iB_{3u}$ state is a possible candidate of the anapole superconductivity. 
Indeed, as discussed above, the BS of the $A_{g}+iB_{3u}$ state exhibits a polar $k_xk_y^2$-type asymmetry, while the BS of the $A_{g}+iA_{u}$ state exhibits a nonpolar $k_xk_yk_z$-type asymmetry (see Fig.~\ref{fig:spectrum}).
It should also be noted that the BS in the $A_{g}+iB_{3u}$ state possesses the polarity along the $k_x$-axis, which coincides with the polar axis of the $B_{3u}^{-}$ IR. 

Based on the above classification and the GL free energy \eqref{eq:GL_free_energy}, the anapole FFLO state with $\bm{q}\propto\bm{T}\parallel\hat{\bm{x}}$ should be naturally realized in the $A_{g}+iB_{3u}$ state. 
In the same manner, we expect the realization of anapole superconducting states with $\bm{T}\parallel\hat{\bm{y}}$ and $\bm{T}\parallel\hat{\bm{z}}$ in the $A_{g}+iB_{2u}$ and $A_{g}+iB_{1u}$ states, respectively (see Supplemental Information for possible anapole superconductivity in UTe$_2$). 

\section*{Discussion}  
From the analogy with magnetic states, we can predict various exotic superconducting phenomena closely related to the asymmetric BS. 
For instance, the asymmetry of the BS will lead to the superconducting analog of the magnetopiezoelectric effect~\cite{HW2017_magnetopiezo,Shiomi2019} and bulk photocurrent response~\cite{HW2021_Photocurrent,Ahn2020}, namely the supercurrent-induced strain and light-induced supercurrent generation, respectively. 
These nonequilibrium phenomena will be useful probes to detect the $\mathcal{P},\mathcal{T}$-symmetry breaking and the asymmetric BS in superconductors. 
Studies for these exotic superconducting phenomena will be presented elsewhere.

\begin{figure}[tbp]
\begin{centering}
\includegraphics[width=1\linewidth]{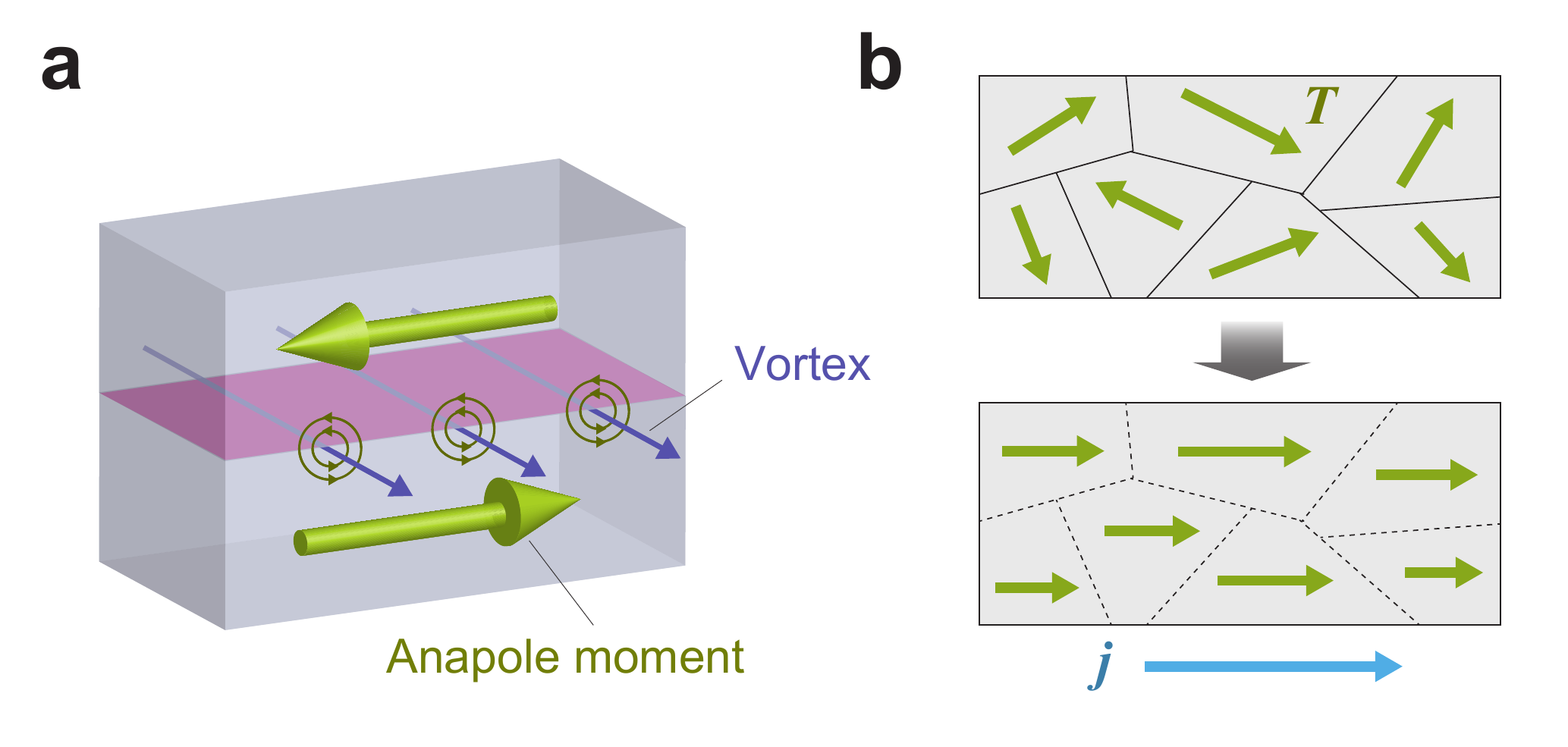}
\end{centering}	
\caption{\textbf{Anapole domain and domain switching through supercurrent.}
(a) Vortices at the boundary of anapole superconducting domains. 
(b) The proposed domain switching in anapole superconductors. The effective anapole moment $\bm{T}$ is aligned along the injected supercurrent $\bm{j}$. }
\label{fig:anapole_domain}
\end{figure}
Experimental detection of the anapole superconductivity should be possible by observing its domain structure. 
The anapole superconducting state effectively carries a supercurrent along the anapole moment $\bm{T}$, since the order parameter is spatially modulated with $e^{i\bm{q}\cdot\bm{r}} \sim e^{i\bm{T}\cdot\bm{r}}$. 
This indicates the emergence of superconducting vortices at the anapole domain boundaries (see Fig.~\ref{fig:anapole_domain}a) even though an external magnetic field is absent. 
Therefore, the observation of vortices at a zero magnetic field can be solid evidence of the anapole superconductivity. 
In addition, the anapole domain can be switched by the supercurrent in a similar way to the electrical switching of antiferromagnets~\cite{Wadley2016,Watanabe2018_DS}.
In an anapole superconductor, the effective anapole moment $\bm{T}$ couples to the applied electric current $\bm{j}$, which is a symmetry-adapted field of the anapole moment. 
Then, the anapole superconducting domain should be switched to align the effective anapole moments along the injected supercurrent $\bm{j}$ (see Fig.~\ref{fig:anapole_domain}b). 
It should also be noticed that the anapole domain switching eliminates the internal magnetic field from the vortices at the domain boundaries, since the domain structure disappears by applying the supercurrent. 
Therefore, the anapole superconducting domain switching can be regarded as a process of erasing magnetic information. 
These properties indicate potential applications of anapole superconductivity as a novel quantum device for magnetic information storage and processing. 

In summary, we have established that the $\mathcal{PT}$-symmetric mixed-parity superconductors generally exhibit asymmetry in the BS. 
The essential ingredient for the asymmetric BS is the $\mathcal{P},\mathcal{T}$-odd nonunitary part of the bilinear product $\Delta\Delta^{\dag}$ arising from the interband pairing. 
Therefore, the multiband nature of superconductivity is essential. 
Especially, we have shown that an FFLO state is stabilized in the absence of an external magnetic field when the superconducting state belongs to a polar and time-reversal-odd IR. 
The stabilization of the FFLO state is evidenced by the emergence of Lifshitz invariants in the free energy due to the effective anapole moment. 
As a specific example, we have shown that the mixed-parity superconductivity in UTe$_2$ can realize the asymmetric BS and anapole superconductivity owing to the locally noncentrosymmetric crystal structure. 
We predicted various superconducting phenomena induced by the asymmetric BS, such as the magnetopiezoelectric effect, nonlinear optical responses, and anapole domain switching from the analogy with magnetic materials.
Topological properties of the anapole superconductivity may also be an intriguing issue. 
Exploration of such exotic phenomena will be a promising route for future research. 

\section*{Methods}
\textbf{Correspondence between Pauli matrices and Dirac matrices.}
In this section, we show that the general form of the BdG Hamiltonian with spin-$1/2$ and a two-valued extra DOF can be expressed by using the Euclidean Dirac matrices. 

Since we assume that the normal state preserves both $\mathcal{P}$- and $\mathcal{T}$-symmetries, $H_0(\bm{k})$ transforms under the space-inversion $\mathcal{P}$ and the time-reversal $\mathcal{T}$ as 
\begin{align}
    H_0(\bm{k}) &\xrightarrow{\mathcal{P}} U_{P}^{\dag}H_0(-\bm{k})U_{P}=H_0(\bm{k}),
    \label{eq:H0_P_trans}
    \\
    H_0(\bm{k}) &\xrightarrow{\mathcal{T}} U_{T}^{\dag}H_0^*(-\bm{k})U_{T}=H_0(\bm{k}),
    \label{eq:H0_T_trans}
\end{align}
where $U_{P}$ and $U_{T}$ are unitary matrices. 
In this paper, we consider a spin-$1/2$ system satisfying $U_TU_T^{*}=-\mathbbm{1}_4$. 
In addition, we require that the time-reversal commute with the space-inversion (i.e., $U_{P}U_{T}=U_{T}U_{P}^{*}$), and the space-inversion operator is its own inverse (i.e., $U_{P}^2=\mathbbm{1}_4$). 
Under the above assumptions, $H_0(\bm{k})$ can be generally expressed as
\begin{align}
    H_{0}(\bm{k}) &= (\epsilon_{\bm{k}}^0-\mu)\sigma_0\otimes\tau_0 + f_{\bm{k}}\sigma_0\otimes\tau_{x_i} 
    \nonumber\\
    & + \bm{g}_{\bm{k}}\cdot\bm{\sigma}\otimes\tau_{y_i}
    + h_{\bm{k}}\sigma_0\otimes\tau_{z_{i}} ,
    \label{eq:H0_spin-orb}
\end{align}
where $\bm{\sigma}=(\sigma_x,\sigma_y,\sigma_z)$ and $\sigma_0\otimes\tau_0=\mathbbm{1}_4$. 
Hermiticity requires all coefficients in Equation \eqref{eq:H0_spin-orb} are real. 
The index $i$ specifies the extra DOF and $(x_i,y_i,z_i)$ is a permutation of $(x, y, z)$. 
Since $U_P$ and $U_T$ vary depending on the extra DOF, the general models \eqref{eq:H0_spin-orb} are classified by the index $i$.
In this paper, we consider three representative examples shown in Table \ref{tab:DOF}.
For $i=1$ ($i=2$), the extra DOF is orbitals with the same (opposite) parity, and $U_{P}=\sigma_0\otimes\tau_{0}$ ($U_{P}=\sigma_0\otimes\tau_{z}$).
For $i=3$, the extra DOF is sublattices in a locally noncentrosymmetric crystal structure, and $U_{P}=\sigma_0\otimes\tau_{x}$.
In these cases, $U_T=i\sigma_y\otimes\tau_{0}$. 
Although the extra DOF can be other than the above three cases, Eq.~\eqref{eq:H0_spin-orb} holds for all the cases unless $U_{P}U_{T}\neq U_{T}U_{P}^{*}$, $U_{P}U_{P}\neq\mathbbm{1}_4$, or $U_{T}U_{T}^{*}\neq-\mathbbm{1}_4$~\cite{Brydon2021AHE}. 
\begin{table}[tbp] 
\caption{\label{tab:DOF}
Classification of two-band models based on the extra DOF. $(x_i,y_i,z_i)$, $U_{P}$, and $U_T$ for $i=1,2,3$ are listed.
}
\centering
\begin{ruledtabular}
{\renewcommand \arraystretch{1.3}
 \begin{tabular}{ccccc} 
& $(x_i,y_i,z_i)$ & $U_{P}$ & $U_{T}$ & DOF
\\ \hline
$i=1$ & $(x,y,z)$ & $\sigma_0\otimes\tau_{0}$ & $i\sigma_y\otimes\tau_{0}$ & orbitals (same parity) \\
$i=2$ & $(z,x,y)$ & $\sigma_0\otimes\tau_{z}$ & $i\sigma_y\otimes\tau_{0}$ & orbitals (opposite parity) \\
$i=3$ & $(y,z,x)$ & $\sigma_0\otimes\tau_{x}$ & $i\sigma_y\otimes\tau_{0}$ & sublattices \\
\end{tabular}
}
\end{ruledtabular}
\end{table} 
Since the set of $\sigma_{\mu}\otimes\tau_{\nu}$ matrices is completely anticommuting in Eq.~\eqref{eq:H0_spin-orb}, we can substitute them by the five anticommuting Euclidean Dirac matrices. 
Then, we can rewrite Equation~\eqref{eq:H0_spin-orb} as Equation~\eqref{eq:H0-gamma}.

The pairing potential $\Delta(\bm{k})$ transforms under the space-inversion and the time-reversal as $\Delta(\bm{k}) \xrightarrow{\mathcal{P}} U_{P}^{\dag}\Delta(-\bm{k})U_{P}^*$ and $\Delta(\bm{k}) \xrightarrow{\mathcal{T}} U_{T}^{\dag}\Delta^*(-\bm{k})U_{T}^*$, respectively. 
In terms of $\hat{\Delta}(\bm{k})=\Delta(\bm{k})U_{T}^{\dag}$, these relations can be rewritten as 
\begin{align}
    \hat{\Delta}(\bm{k}) &\xrightarrow{\mathcal{P}} U_{P}^{\dag}\hat{\Delta}(-\bm{k})U_{P}, \label{eq:Delta_P_trans_hat}
    \\
    \hat{\Delta}(\bm{k}) &\xrightarrow{\mathcal{T}} \hat{\Delta}^{\dag}(\bm{k}).
    \label{eq:Delta_T_trans_hat}
\end{align}
We note that Equation~\eqref{eq:Delta_P_trans_hat} is equivalent to the transformation property of $H_0(\bm{k})$ under the space-inversion [see Eq.~\eqref{eq:H0_P_trans}], while Eq.~\eqref{eq:Delta_T_trans_hat} corresponds to the Hermiticity condition. 
Whereas $H_0(\bm{k})$ is assumed to preserve both $\mathcal{P}$- and $\mathcal{T}$-symmetries, we admit that $\Delta(\bm{k})$ spontaneously breaks the $\mathcal{P}$- and $\mathcal{T}$-symmetries. 
The only requirements for the pairing potential is satisfying the fermionic antisymmetry $\Delta(\bm{k}) = -\Delta^{\rm T}(-\bm{k})$, which can be rewritten as
\begin{align}
    \hat{\Delta}(\bm{k}) &= U_{T}^{\dag}\hat{\Delta}^{\rm T}(-\bm{k})U_{T},
    \label{eq:fermionic_antisymmetry_hat}
\end{align}
where we used the fact that $U_{T}^{\dag}=U_{T}^{\rm T}=-U_{T}$ by choosing $U_T$ as real (i.e., $U_T=U_T^*$). 
It should be noticed that Eq.~\eqref{eq:fermionic_antisymmetry_hat} is formally equivalent to the time-reversal symmetry for $H_0(\bm{k})$ [see Eq.~\eqref{eq:H0_T_trans}].
Since the even-parity part of $\hat{\Delta}(\bm{k})$ obeys transformation properties completely equivalent to those of $H_0(\bm{k})$ under the time-reversal and the space-inversion, it can be expressed as a linear combination of six $\sigma_{\mu}\otimes\tau_{\nu}$ matrices allowed to appear in $H_0(\bm{k})$.
On the other hand, the other ten $\sigma_{\mu}\otimes\tau_{\nu}$ matrices, which correspond to $i\gamma_m\gamma_n$ ($1\leq m<n \leq 5$), constitute the odd-parity pairing potential. 
Then, we obtain a general form of $\Delta(\bm{k})$ as
\begin{align}
    \hat{\Delta}(\bm{k}) &= \Delta_{1}\left[\sum_{\nu=0,x_i,z_i}\psi^{\nu}_{\bm{k}}\sigma_0\otimes\tau_{\nu} + \bm{d}^{y_i}_{\bm{k}}\cdot\bm{\sigma}\otimes\tau_{y_i}\right]
    \nonumber \\
    &+ \Delta_{2}\left[\sum_{\nu=0,x_i,z_i}\bm{d}^{\nu}_{\bm{k}}\cdot\bm{\sigma}\otimes\tau_{\nu} + \psi^{y_i}_{\bm{k}}\sigma_0\otimes\tau_{y_i} \right] ,
    \label{eq:Delta_PauliForm}
\end{align}
where $\psi^{\nu}_{\bm{k}}$ and $\bm{d}^{\nu}_{\bm{k}}$ are real-valued coefficients. 
Note that $\Delta_{1}$ and $\Delta_2$ are complex valued since $\hat{\Delta}(\bm{k})\neq\hat{\Delta}^{\dag}(\bm{k})$ in $\mathcal{T}$-symmetry breaking superconducting phases.
From Equation~\eqref{eq:Delta_PauliForm}, we obtain Equation~\eqref{eq:Delta-gamma} as a general form of $\Delta(\bm{k})$ in two-band models. 

\begin{table}[tbp] 
\caption{\label{tab:conditions_Pauli}
Necessary conditions for the asymmetric BS in two-band models ($\sigma_{\mu}\otimes\tau_{\nu}$ basis). 
}
\centering
\begin{ruledtabular}
{\renewcommand \arraystretch{1.3}
 \begin{tabular}{ccc} 
   & Criterion (i) & Criterion (ii)
  \\ \hline
  (I) &  & $\psi_{\bm{k}}^{z_i}\psi_{\bm{k}}^{y_i}f_{\bm{k}}\neq0$
  \\
  (II) & & $(\bm{d}_{\bm{k}}^{y_i}\cdot\bm{d}_{\bm{k}}^{z_i})f_{\bm{k}}\neq0$
  \\
  (III) & & $\psi_{\bm{k}}^{x_i}(\bm{d}_{\bm{k}}^{z_i}\cdot\bm{g}_{\bm{k}})\neq0$
  \\ 
  (IV) & $\mathrm{Im}(\Delta_1\Delta_2^{*})\neq0$ & $\psi_{\bm{k}}^{z_i}(\bm{d}_{\bm{k}}^{x_i}\cdot\bm{g}_{\bm{k}})\neq0$
  \\ 
  (V) & & $(\bm{d}_{\bm{k}}^{y_i}\times\bm{d}_{\bm{k}}^{0})\cdot\bm{g}_{\bm{k}}\neq0$
  \\ 
  (VI) & & $(\bm{d}_{\bm{k}}^{y_i}\cdot\bm{d}_{\bm{k}}^{x_i})h_{\bm{k}}\neq0$
  \\
  (VII) & & $\psi_{\bm{k}}^{x_i}\psi_{\bm{k}}^{y_i}h_{\bm{k}}\neq0$
\end{tabular}
}
\end{ruledtabular}
\end{table}
From Equations~\eqref{eq:H0_spin-orb} and \eqref{eq:Delta_PauliForm}, we obtain
\begin{align}
    \mathrm{Tr}[M_{-}^{(1)}(\bm{k})\tilde{H}_0(\bm{k})] 
    &= 8\mathrm{Im}(\Delta_1\Delta_2^*) \times
    \nonumber \\
    & \big[ 
    (\psi_{\bm{k}}^{z_i}\psi_{\bm{k}}^{y_i}-\bm{d}_{\bm{k}}^{y_i}\cdot\bm{d}_{\bm{k}}^{z_i})f_{\bm{k}} 
    \nonumber \\
    &+ (\psi_{\bm{k}}^{x_i}\bm{d}_{\bm{k}}^{z_i}-\psi_{\bm{k}}^{z_i}\bm{d}_{\bm{k}}^{x_i}-\bm{d}_{\bm{k}}^{y_i}\times\bm{d}_{\bm{k}}^{0})\cdot\bm{g}_{\bm{k}}
    \nonumber \\
    &+ (\bm{d}_{\bm{k}}^{y_i}\cdot\bm{d}_{\bm{k}}^{x_i}-\psi_{\bm{k}}^{x_i}\psi_{\bm{k}}^{y_i})h_{\bm{k}}
    \big].
\end{align}
Then, in the $\sigma_{\mu}\otimes\tau_{\nu}$ basis, the necessary conditions for the asymmetric BS (i.e., $\mathrm{Tr}[M_{-}^{(1)}(\bm{k})\tilde{H}_{0}(\bm{k})]\neq0$) can be summarized as shown in Table \ref{tab:conditions_Pauli}. 
For example, the condition (I) means that the asymmetric BS appears when $\mathrm{Im}(\Delta_1\Delta_2^{*})\neq0$ and $\psi_{\bm{k}}^{z_i}\psi_{\bm{k}}^{y_i}f_{\bm{k}}\neq0$. \\


\textbf{Asymmetry of BS in the minimal two-band model. }
We here prove that Equation~\eqref{eq:spectrum} indeed expresses the asymmetric BS. 
For $r=1$, Equation~\eqref{eq:spectrum} leads to
\begin{align}
   E_{\bm{k}}^{\pm}=&\sqrt{\xi_{\bm{k}}^2+|\Delta_1\eta_{\bm{k}}^{b}|^2+|\Delta_2\eta_{\bm{k}}^{ab}|^2\pm 2\mathrm{Im}(\Delta_1\Delta_2^*)\eta_{\bm{k}}^b\eta_{\bm{k}}^{ab}}
   \nonumber\\
   &\pm\epsilon_{\bm{k}}^{a} .
   \label{eq:spectrum-1}
\end{align} 
Then, we need to specify the $\bm{k}$-parity of $\epsilon_{\bm{k}}^a$, $\eta_{\bm{k}}^b$, and $\eta_{\bm{k}}^{ab}$, which depend on the details of the extra DOF, to investigate the property of the BS $E_{-\bm{k}}^{\pm}$. 
We here denote $\epsilon_{-\bm{k}}^a=p_a\epsilon_{\bm{k}}^a$, $\eta_{-\bm{k}}^b=p_b\eta_{\bm{k}}^b$, and $\eta_{-\bm{k}}^{ab}=p_{ab}\eta_{\bm{k}}^{ab}$ ($p_a,p_b,p_{ab}=\pm1$). 
From Equations~\eqref{eq:H0_P_trans} and \eqref{eq:H0_T_trans}, we obtain $p_{a}\gamma_a = U_{T}^{\dag}\gamma_a^*U_T = U_{P}^{\dag}\gamma_aU_P$. 
On the other hand, the $\mathcal{P,T}$-odd behavior of $M_{-}^{(1)}(\bm{k})=2\mathrm{Im}(\Delta_1\Delta_2^*)\eta_{\bm{k}}^b\eta_{\bm{k}}^{ab}\gamma_a$ leads to $-p_{b}p_{ab}\gamma_a = U_{T}^{\dag}\gamma_a^*U_T = U_{P}^{\dag}\gamma_aU_P$. 
Thus, $p_{a}=-p_{b}p_{ab}$ holds in general. 
Using this relation, we obtain 
\begin{align}
   E_{-\bm{k}}^{\pm}=&\sqrt{\xi_{\bm{k}}^2+|\Delta_1\eta_{\bm{k}}^{b}|^2+|\Delta_2\eta_{\bm{k}}^{ab}|^2\mp p_a 2\mathrm{Im}(\Delta_1\Delta_2^*)\eta_{\bm{k}}^b\eta_{\bm{k}}^{ab}}
   \nonumber\\
   &\pm p_a\epsilon_{\bm{k}}^{a} \quad (p_a=\pm1).
   \label{eq:spectrum-2}
\end{align} 
Comparing Equation~\eqref{eq:spectrum-2} with \eqref{eq:spectrum-1}, we can safely say that $E_{-\bm{k}}^{\pm}\neq E_{\bm{k}}^{+},E_{\bm{k}}^{-}$ and the BS is asymmetric. 
In the same manner, we can prove the asymmetry of Equation~\eqref{eq:spectrum} for $r=0$. \\

\textbf{Data Availability Statement}\\
The data that support the findings of this study are available from the corresponding author upon reasonable request. \\

\textbf{Acknowledgements}\\
The authors are grateful to Jun Ishizuka, Hikaru Watanabe, and Shuntaro Sumita for helpful discussions. 
This work was supported by JSPS KAKENHI (Grants No. JP18H05227, No. JP18H01178, and No. JP20H05159) and by SPIRITS 2020 of Kyoto University.
S.K. is supported by a JSPS research fellowship and by JSPS KAKENHI (Grant No. 19J22122). \\

\textbf{Author Contributions}\\
S.K. and Y.Y. conceived the idea and initiated the project. 
S.K. performed the major part of the calculations. 
S.K. and Y.Y. discussed the results and co-wrote the paper. \\

\textbf{Conflict of interest statement}\\
The authors declare no competing interests.



\providecommand{\noopsort}[1]{}\providecommand{\singleletter}[1]{#1}%
%

\clearpage

\renewcommand{\thesection}{S\arabic{section}}
\renewcommand{\theequation}{S\arabic{equation}}
\setcounter{equation}{0}
\renewcommand{\thefigure}{S\arabic{figure}}
\setcounter{figure}{0}
\renewcommand{\thetable}{S\arabic{table}}
\setcounter{table}{0}
\renewcommand*{\citenumfont}[1]{S#1}
\renewcommand*{\bibnumfmt}[1]{[S#1]}
\makeatletter
\c@secnumdepth = 2
\makeatother

\onecolumngrid
\begin{center}
{\large \textmd{Supplemental Information} \\[0.3em]
{\bfseries Anapole superconductivity from $\mathcal{PT}$-symmetric mixed-parity interband pairing} \\[0.9em]
Shota Kanasugi and Youichi Yanase}
\end{center}

\setcounter{page}{1}

\section{Parity- and time-reversal-odd bilinear product}
In this section, we derive a formula to calculate the parity- and time-reversal-odd bilinear product, which is used in the main text.
To obtain the formula, we first consider the transformation property of the bilinear product $\Delta(\bm{k})\Delta^{\dag}(\bm{k})$ under the time-reversal. 
Since the time-reversed counterpart of $\hat{\Delta}(\bm{k})=\Delta(\bm{k})U_T^{\dag}$ is $\hat{\Delta}^{\dag}(\bm{k})$, the transformation property of the bilinear product $\Delta(\bm{k})\Delta^{\dag}(\bm{k})=\hat{\Delta}(\bm{k})\hat{\Delta}^{\dag}(\bm{k})$ under the time-reversal $\mathcal{T}$ is obtained as
\begin{align}
    \Delta(\bm{k})\Delta^{\dag}(\bm{k})=\hat{\Delta}(\bm{k})\hat{\Delta}^{\dag}(\bm{k}) &\xrightarrow{\mathcal{T}}   \hat{\Delta}^{\dag}(\bm{k})\hat{\Delta}(\bm{k}).
\end{align}
Then, we define the time-reversal-odd bilinear product $M^{(1)}(\bm{k})$ as 
\begin{equation}
    M^{(1)}(\bm{k}) = \frac{1}{2}\left[\hat{\Delta}(\bm{k})\hat{\Delta}^{\dag}(\bm{k})-\hat{\Delta}^{\dag}(\bm{k})\hat{\Delta}(\bm{k})\right] 
    = \frac{1}{2}[\hat{\Delta}(\bm{k}),\hat{\Delta}^{\dag}(\bm{k})].
    \label{eq:T_odd_gap_prod}
\end{equation}
Equation~\eqref{eq:T_odd_gap_prod} extracts the time-reversal-odd part of the bilinear product $\Delta(\bm{k})\Delta^{\dag}(\bm{k})$ \cite{S_Brydon2019AHE,S_Brydon2021AHE}. 
Here, we decompose the pairing potential $\Delta(\bm{k})$ into the even-parity part $\Delta^{g}(\bm{k})$ and odd-parity part $\Delta^{u}(\bm{k})$ as
\begin{align}
    \Delta(\bm{k})=\Delta^{g}(\bm{k})+\Delta^{u}(\bm{k}).
\end{align}
Then, $\Delta(\bm{k})$ transforms under the space-inversion $\mathcal{P}$ as 
\begin{align}
    \Delta(\bm{k})=\Delta^{g}(\bm{k})+\Delta^{u}(\bm{k}) &\xrightarrow{\mathcal{P}} \Delta^{g}(\bm{k})-\Delta^{u}(\bm{k}).
    \label{eq:Delta_P_trans_mixed}
\end{align}
From Eq.~\eqref{eq:Delta_P_trans_mixed}, $M^{(1)}(\bm{k})$ transforms under the space-inversion $\mathcal{P}$ as
\begin{align}
    M^{(1)}(\bm{k}) = M_{+}^{(1)}(\bm{k})+ M_{-}^{(1)}(\bm{k}) \xrightarrow{\mathcal{P}} M_{+}^{(1)}(\bm{k}) - M_{-}^{(1)}(\bm{k}),
\end{align}
where
\begin{align}
    M_{+}^{(1)}(\bm{k}) &= \frac{1}{2}\left([\hat{\Delta}^{g}(\bm{k}),\hat{\Delta}^{g\dag}(\bm{k})]+[\hat{\Delta}^{u}(\bm{k}),\hat{\Delta}^{u\dag}(\bm{k})]\right),
    \label{eq:M_plus}
    \\
    M_{-}^{(1)}(\bm{k}) &= \frac{1}{2}\left([\hat{\Delta}^{g}(\bm{k}),\hat{\Delta}^{u\dag}(\bm{k})]+[\hat{\Delta}^{u}(\bm{k}),\hat{\Delta}^{g\dag}(\bm{k})]\right).
    \label{eq:M_minus}
\end{align}
$M_{+}^{(1)}(\bm{k})$ and $M_{-}^{(1)}(\bm{k})$ are the even-parity and odd-parity part of the time-reversal-odd bilinear product $M^{(1)}(\bm{k})$, respectively. 
Then, Eq.~\eqref{eq:M_minus} represents the parity- and time-reversal-odd nonunitary part of $\Delta(\bm{k})\Delta^{\dag}(\bm{k})$. 

\section{Ginzburg–Landau Free energy}
In this section, we perform the Ginzburg–Landau (GL) expansion of the free energy for the mixed-parity superconductivity. 
Then, we derive an analytical expression of the GL free energy for a model satisfying one of the necessary conditions to realize the asymmetric Bogoliubov spectrum (BS), which is derived in the main text. 

We consider the Hamiltonian $\mathcal{H}=\mathcal{H}_0+\mathcal{H}_{\rm int}$, which is composed of the single-particle term $\mathcal{H}_0$ and pairing interaction term $\mathcal{H}_{\rm int}$. 
The pairing interaction $\mathcal{H}_{\rm int}$ is assumed to be a mixture of even-parity and odd-parity channels as
\begin{align}
    \mathcal{H}_{\rm int} &= \frac{1}{2}\sum_{j=1,2}V_{j}B_{j}^{\dag}(\bm{q})B_{j}(\bm{q}), 
    \label{eq:Hint}
\end{align}
where $\bm{q}=(q_x,q_y,q_z)$ is the center-of-mass momentum of the Cooper pairs, $V_{j}(<0)$ is the strength of the pairing interaction, and $j=1$ ($j=2$) represents an index of the even-parity (odd-parity) pairing channel. 
Note that we assume a single-$\bm{q}$ state in Eq.~\eqref{eq:Hint}. 
The creation operator of the Cooper pairs $B_{j}^{\dag}(\bm{q})$ is given by
\begin{align}
    B_{j}^{\dag}(\bm{q}) &= \sum_{\bm{k}}\sum_{ll',ss'}\varphi_{ls,l's'}^{j}(\bm{k})c_{\bm{k}+\bm{q}/2,ls}^{\dag}c_{-\bm{k}+\bm{q}/2,l's'}^{\dag}, 
\end{align}
where $s,s'= \uparrow,\downarrow$ and $l,l'= 1,2$ are indexes for the spin-$1/2$ and extra two-valued DOF, respectively. 
Here, we apply the mean-field approximation to $\mathcal{H}_{\rm int}$ as
\begin{equation}
    \mathcal{H}_{\rm int} \approx \frac{1}{2}\sum_{j=1,2}\sum_{\bm{k}}\sum_{ls,l's'}\left[\Delta_{j}\varphi_{ls,l's'}^{j}(\bm{k})c_{\bm{k}+\bm{q}/2,ls}^{\dag}c_{-\bm{k}+\bm{q}/2,l's'}^{\dag}+\mathrm{H.c.}\right] - \sum_{j=1,2}\frac{|\Delta_{j}|^2}{V_{j}},
    \label{eq:Hint_MF}
\end{equation}
by introducing the superconducting order parameter
\begin{equation}
    \Delta_{j} = V_{j}\sum_{\bm{k}}\sum_{ll',ss'}\varphi_{ls,l's'}^{j\dag}(\bm{k})\average{c_{-\bm{k}+\bm{q}/2,ls}c_{\bm{k}+\bm{q}/2,l's'}}.
\end{equation}
Then, a matrix form of the total Hamiltonian $\mathcal{H}$ is obtained as
\begin{align}
   \mathcal{H} = \frac{1}{2}\sum_{\bm{k}}
      (\hat{c}_{\bm{k}}^{\dag} , \hat{c}_{-\bm{k}}^{\rm T} )
    \begin{pmatrix}
      H_0(\bm{k}+\bm{q}/2) & \Delta(\bm{k}) \\
       \Delta^{\dag}(\bm{k}) & -H_0^{\rm T}(-\bm{k}+\bm{q}/2)
    \end{pmatrix}
    \begin{pmatrix}
      \hat{c}_{\bm{k}} \\
      \hat{c}_{-\bm{k}}^{\dag}
    \end{pmatrix}
    - \sum_{j=1,2}\frac{|\Delta_{j}|^2}{V_{j}},
    \label{eq:H_mat}
\end{align}
where $\hat{c}_{\bm{k}}^{\rm T}=(c_{\bm{k}1\uparrow},c_{\bm{k}1\downarrow},c_{\bm{k}2\uparrow},c_{\bm{k}2\downarrow})$ and some constants are omitted in Eq.~\eqref{eq:H_mat}. 
The pairing potential $\Delta(\bm{k})$ is given by
\begin{align}
    \Delta_{ls,l's'}(\bm{k}) = \sum_{j=1,2} \Delta_j \varphi_{ls,l's'}^{j}(\bm{k}). 
\end{align}

To obtain the GL free energy for the $\mathcal{PT}$-symmetric mixed-parity superconducting states with the asymmetric BS, we here assume that the pairing potentials are described as
\begin{align}
    \varphi^{1}(\bm{k}) &= [r\eta_{\bm{k}}^{b}\gamma_{b}+(1-r)\eta_{\bm{k}}^{a}\gamma_{a}]U_{T},
    \\
    \varphi^{2}(\bm{k}) &= \eta_{\bm{k}}^{ab}i\gamma_{a}\gamma_{b}U_{T},
\end{align}
where $a$ and $b$ are integers satisfying $1\leq a< b\leq5$ and $r$ takes the value either 0 or 1. 
In addition, we suppose that the normal state Hamiltonian $H_0(\bm{k})$ is described as 
\begin{align}
    H_0(\bm{k}) = \xi_{\bm{k}}\mathbbm{1}_{4} + r\epsilon_{\bm{k}}^{a}\gamma_{a}+(1-r)\epsilon_{\bm{k}}^{b}\gamma_{b}. 
\end{align}
Then, the model satisfies one of the necessary conditions for the asymmetric BS, which is shown in the main text. 
By assuming $|\xi_{\bm{k}}|\gg\mathrm{max}(|\epsilon_{\bm{k}}^{a,b}|)$, we can approximate Eq.~\eqref{eq:H_mat} as 
\begin{align}
    \mathcal{H} \approx \frac{1}{2}\sum_{\bm{k}}
      (\hat{c}_{\bm{k}}^{\dag} , \hat{c}_{-\bm{k}}^{\rm T} )
    \begin{pmatrix}
      H_0(\bm{k})+\frac{1}{2}\bm{v}_{\bm{k}}\cdot\bm{q}\mathbbm{1}_4 & \Delta(\bm{k}) \\
       \Delta^{\dag}(\bm{k}) & -H_0^{\rm T}(-\bm{k})+\frac{1}{2}\bm{v}_{\bm{k}}\cdot\bm{q}\mathbbm{1}_4
    \end{pmatrix}
    \begin{pmatrix}
      \hat{c}_{\bm{k}} \\
      \hat{c}_{-\bm{k}}^{\dag}
    \end{pmatrix}
    - \sum_{j=1,2}\frac{|\Delta_{j}|^2}{V_{j}} ,
    \label{eq:H_mat_approx}
\end{align}
where $\bm{v}_{\bm{k}}=(v_{\bm{k}}^x,v_{\bm{k}}^y,v_{\bm{k}}^z)\equiv\nabla\xi_{\bm{k}}$ is the Fermi velocity. 
By diagonalizing the BdG Hamiltonian matrix in Eq.~\eqref{eq:H_mat_approx}, we can obtain the free energy $\mathcal{F}$ as follows: 
\begin{align}
    \mathcal{F} = -\frac{2}{\beta}\sum_{\bm{k}}\sum_{\sigma=\pm}\left[\ln{\left(1+e^{-\beta(E_{\bm{k}}^{\sigma}+\bm{v}_{\bm{k}}\cdot\bm{q}/2)}\right)} + \ln{\left(1+e^{-\beta(-E_{-\bm{k}}^{\sigma}+\bm{v}_{\bm{k}}\cdot\bm{q}/2)}\right)}\right] - \sum_{j=1,2}\frac{|\Delta_{j}|^2}{V_{j}},
    \label{eq:F}
\end{align}
where $\beta=1/T$ is the inverse temperature. 
The quasiparticle energy $E_{\bm{k}}^{\pm}$ is given by
\begin{equation}
    E_{\bm{k}}^{\pm} = \sqrt{\xi_{\bm{k}}^2+\frac{1}{4}\mathrm{Tr}\left[\Delta(\bm{k})\Delta^{\dag}(\bm{k})\pm\frac{M_{-}^{(1)}(\bm{k})\tilde{H}_0(\bm{k})}{r\epsilon_{\bm{k}}^{a}+(1-r)\epsilon_{\bm{k}}^{b}}\right]} \pm [r\epsilon_{\bm{k}}^{a}+(1-r)\epsilon_{\bm{k}}^{b}].
    \label{eq:Bogo_spect}
\end{equation}
By differentiating Eq.~\eqref{eq:F} with respect to $\Delta_j$ and $\Delta_j^*$, we obtain an analytical expression of the GL free energy as 
\begin{align}
    \mathcal{F} &= \alpha_1|\Delta_{1}|^2+\alpha_2|\Delta_{2}|^2 + \beta_{1}|\Delta_{1}|^4 + \beta_{2}|\Delta_{2}|^4 
    + 4\tilde{\beta}|\Delta_{1}|^2|\Delta_{2}|^2 - \tilde{\beta}(\Delta_{1}^{2}\Delta_{2}^{*2}+\Delta_{2}^{2}\Delta_{1}^{*2})
    \nonumber\\
    & \quad + \sum_{\nu=x,y,z}(\kappa_{1,\nu}|\Delta_{1}|^{2} + \kappa_{2,\nu}|\Delta_{2}|^{2} )q_{\nu}^2 + \bm{T}\cdot\bm{q}  .
    \label{eq:GL_free_energy}
\end{align}
The coefficients of the quadratic terms are given by
\begin{align}
    \alpha_{1} &= \frac{1}{|V_1|} - 2\sum_{\bm{k}}|\bm{\eta}_{\bm{k}}|^2\frac{1-2f(|\xi_{\bm{k}}|)}{|\xi_{\bm{k}}|}
    \approx \rho_0\average{|\bm{\eta}_{\bm{k}}|^2}_{\rm FS}\frac{T-T_{{\rm c},1}}{T_{{\rm c},1}}, 
    \label{eq:alpha1}
    \\
    \alpha_{2} &= \frac{1}{|V_2|} - 2\sum_{\bm{k}}|\eta_{\bm{k}}^{ab}|^2\frac{1-2f(|\xi_{\bm{k}}|)}{|\xi_{\bm{k}}|} 
    \approx \rho_0\average{|\eta^{ab}_{\bm{k}}|^2}_{\rm FS}\frac{T-T_{{\rm c},2}}{T_{{\rm c},2}},
    \label{eq:alpha2}
\end{align}
where $|\bm{\eta}_{\bm{k}}|^2\equiv[r|\eta_{\bm{k}}^{b}|^2+(1-r)|\eta_{\bm{k}}^{a}|^2]$, $f(x)=1/(e^{\beta x}+1)$ is the Fermi-Dirac distribution function, $\rho_0$ is the density of states at the Fermi energy, and $\average{\cdots}_{\rm FS}$ denotes the average over the Fermi surface. 
The superconducting transition temperature for the even-parity and odd-parity pairing channel $T_{{\rm c},1}$ and $T_{{\rm c},2}$ are defined as
\begin{align}
    T_{{\rm c},1}&=\frac{2e^{\gamma}}{\pi}\epsilon_c\exp{\left(-\frac{1}{\rho_0\average{|\bm{\eta}_{\bm{k}}|^2}_{\rm FS}|V_1|}\right)},
    \\
    T_{{\rm c},2}&=\frac{2e^{\gamma}}{\pi}\epsilon_c\exp{\left(-\frac{1}{\rho_0\average{|\eta_{\bm{k}}^{ab}|^2}_{\rm FS}|V_2|}\right)},
\end{align}
where $\gamma=0.577\cdots$ is the Euler's constant, and $\epsilon_c$ is a cutoff energy. 
In Eqs.~\eqref{eq:alpha1} and \eqref{eq:alpha2}, the summation over $\bm{k}$ is approximated as
\begin{align}
    \sum_{\bm{k}}X(\bm{k})Y(\xi_{\bm{k}}) &\approx \int_{\rm FS}\frac{dk_{\rm F}}{v(\bm{k}_{\rm F})}X(\bm{k}_{\rm F})\int_{-\epsilon_c}^{\epsilon_c}d\xi Y(\xi) 
    \approx \frac{\rho_0}{4}\average{X(\bm{k}_{\rm F})}_{\rm FS}\int_{-\epsilon_c}^{\epsilon_c}d\xi Y(\xi) , 
    \label{eq:approx_sum_k}
\end{align}
where $X$ and $Y$ are some functions, and $\bm{k}_{\rm F}$ is the Fermi wave vector. 
The coefficients of the quartic terms are given by
\begin{align}
    \beta_{1} &= \frac{1}{2}\sum_{\bm{k}}|\bm{\eta}_{\bm{k}}|^4\left[\frac{1-2f(|\xi_{\bm{k}}|)}{|\xi_{\bm{k}}|^3} + \frac{2f'(|\xi_{\bm{k}}|)}{|\xi_{\bm{k}}|^2} \right] 
    \approx \rho_0\average{|\bm{\eta}_{\bm{k}}|^4}_{\rm FS}\frac{7\zeta(3)}{16\pi^2T^2},
    \label{eq:beta1}
    \\
    \beta_{2} &= \frac{1}{2}\sum_{\bm{k}}|\eta^{ab}_{\bm{k}}|^4\left[\frac{1-2f(|\xi_{\bm{k}}|)}{|\xi_{\bm{k}}|^3} + \frac{2f'(|\xi_{\bm{k}}|)}{|\xi_{\bm{k}}|^2} \right] 
    \approx \rho_0\average{|\eta^{ab}_{\bm{k}}|^4}_{\rm FS}\frac{7\zeta(3)}{16\pi^2T^2},
    \label{eq:beta2}
    \\
    \tilde{\beta} &= \frac{1}{2}\sum_{\bm{k}}|\bm{\eta}_{\bm{k}}|^2|\eta^{ab}_{\bm{k}}|^2\left[\frac{1-2f(|\xi_{\bm{k}}|)}{|\xi_{\bm{k}}|^3} + \frac{2f'(|\xi_{\bm{k}}|)}{|\xi_{\bm{k}}|^2} \right] 
    \approx \rho_0\average{|\bm{\eta}_{\bm{k}}|^2|\eta^{ab}_{\bm{k}}|^2}_{\rm FS}\frac{7\zeta(3)}{16\pi^2T^2},
    \label{eq:beta12}
\end{align}
where $f'(x)=df(x)/dx$, and $\zeta(x)$ is the Riemann zeta function. In Eqs.~\eqref{eq:beta1}-\eqref{eq:beta12}, we used the following integral formula;
\begin{align}
    \int_{-\infty}^{\infty}d\xi\left[\frac{1-2f(\xi)}{\xi^3} + \frac{2f'(\xi)}{\xi^2}\right] = \int_{-\infty}^{\infty}d\xi\frac{f''(\xi)}{\xi} = \frac{7\zeta(3)}{2(\pi T)^2} ,
    \label{eq:integral}
\end{align}
where $f''(x)=df'(x)/dx$. 
The coefficients of the quadratic gradient term are obtained as
\begin{align}
    \kappa_{1,\nu} &= \frac{1}{2}\sum_{\bm{k}}|\bm{\eta}_{\bm{k}}|^2|v^{\nu}_{\bm{k}}|^2\frac{f''(|\xi_{\bm{k}}|)}{|\xi_{\bm{k}}|}
    \approx \rho_0\average{|\bm{\eta}_{\bm{k}}|^2|v^{\nu}_{\bm{k}}|^2}_{\rm FS}\frac{7\zeta(3)}{16\pi^2T^2},
    \\
    \kappa_{2,\nu} &= 
    \frac{1}{2}\sum_{\bm{k}}|\eta^{ab}_{\bm{k}}|^2|v^{\nu}_{\bm{k}}|^2\frac{f''(|\xi_{\bm{k}}|)}{|\xi_{\bm{k}}|}
    \approx \rho_0\average{|\eta^{ab}_{\bm{k}}|^2|v^{\nu}_{\bm{k}}|^2}_{\rm FS}\frac{7\zeta(3)}{16\pi^2T^2}.
\end{align}
where we used Eq.~\eqref{eq:integral}. 
In the same manner, the effective anapole moment $\bm{T}$ is given by
\begin{align}
    \bm{T} &= \frac{1}{2}\sum_{\bm{k}}\mathrm{Tr}[M_{-}^{(1)}(\bm{k})\tilde{H}_{0}(\bm{k})]\bm{v}_{\bm{k}}\frac{f''(|\xi_{\bm{k}}|)}{|\xi_{\bm{k}}|}
    \approx \rho_0\average{\mathrm{Tr}[M_{-}^{(1)}(\bm{k})\tilde{H}_{0}(\bm{k})]\bm{v}_{\bm{k}}}_{\rm FS}\frac{7\zeta(3)}{16\pi^2T^2}. 
\end{align}

\section{Symmetry analysis for $\mathrm{UTe}_{2}$}
In this section, we present a symmetry analysis for possible asymmetric BS and anapole superconductivity in UTe$_2$. 
Although we considered only the $A_g+iA_u$ and $A_g+iB_{3u}$ states in the main text, we here consider all of possible $\mathcal{PT}$-symmetric mixed-parity pairing states in UTe$_2$. 
The superconducting order parameter in UTe$_2$ is classified based on the eight irreducible representations (IRs) in $D_{2h}$ point group. 
The basis functions for these pairing states are shown in Table \ref{tab:UTe2_basis}.
Since the local site symmetry at U site is $C_{2v}$ in UTe$_2$ \cite{S_Ishizuka2020}, the basis functions for the staggered pairing components can be obtained as listed in the third column of Table \ref{tab:UTe2_basis}.
As shown in the main text, the staggered pairing components and antisymmetric spin-orbit coupling are essential for the asymmetric BS. 

There are 16 patterns of $\mathcal{PT}$-symmetric mixed-parity pairing as shown in Table \ref{tab:UTe2}. 
If the pairing state belongs to the nonpolar $A_u^{-}$ IR, a nonpolar $k_xk_yk_z$-type asymmetry can be induced in the BS. The $A_u^{-}$ pairing states are equivalent to nonpolar odd-parity magnetic multipole states such as magnetic monopole, quadrupole, and hexadecapole, from the viewpoint of symmetry. 
On the other hand, if the pairing state belongs to the polar $B_{1u,2u,3u}^{-}$ IRs, the BS can exhibit a polar $k_{z,y,x}$-type asymmetry. 
Thus, the $B_{1u,2u,3u}^{-}$ pairing state carries the anapole (magnetic toroidal) moment.
This $k_{\nu}$-type asymmetry leads to stabilization of Fulde-Ferrell-Larkin-Ovchinnikov (FFLO) state with $q_{\nu}\neq0$ ($\nu=x,y,z$), where $\bm{q}=(q_x,q_y,q_z)$ is the center-of-mass momentum of Cooper pairs.  
\begin{table}[htbp]
 \centering
 \caption{Basis functions for possible superconducting states in UTe$_2$. The second (third) column shows the basis function for intrasublattice (staggered) pairing components, which is proportional to $\tau_0$ ($\tau_z$). Here, $\tau_{\nu}$ is the Pauli matrices for intra-ladder sublattice DOF in UTe$_2$ \cite{S_Ishizuka2020}. }
 \label{tab:UTe2_basis}
 \begin{ruledtabular}
{\renewcommand \arraystretch{1.3}
 \begin{tabular}{ccc}
  IR & Intrasublattice components ($\sim\tau_0$) & Staggered components ($\sim\tau_z$) \\ \hline
  $A_{g}$ & $1$ & $k_y\hat{\bm{x}}$, $k_x\hat{\bm{y}}$, $k_xk_yk_z\hat{\bm{z}}$ \\
  $B_{1g}$ & $k_xk_y$ & $k_x\hat{\bm{x}}$, $k_y\hat{\bm{y}}$, $k_z\hat{\bm{z}}$ \\
  $B_{2g}$ & $k_zk_x$ & $k_xk_yk_z\hat{\bm{x}}$, $k_z\hat{\bm{y}}$, $k_y\hat{\bm{z}}$ \\
  $B_{3g}$ & $k_yk_z$ & $k_z\hat{\bm{x}}$, $k_xk_yk_z\hat{\bm{y}}$, $k_x\hat{\bm{z}}$ \\
  $A_{u}$ & $k_x\hat{\bm{x}}$, $k_y\hat{\bm{y}}$, $k_z\hat{\bm{z}}$ & $k_xk_y$ \\
  $B_{1u}$ & $k_y\hat{\bm{x}}$, $k_x\hat{\bm{y}}$, $k_xk_yk_z\hat{\bm{z}}$ & $1$ \\
  $B_{2u}$ & $k_z\hat{\bm{x}}$, $k_xk_yk_z\hat{\bm{y}}$, $k_x\hat{\bm{z}}$ & $k_yk_z$ \\
  $B_{3u}$ & $k_xk_yk_z\hat{\bm{x}}$, $k_z\hat{\bm{y}}$, $k_y\hat{\bm{z}}$ & $k_zk_x$ \\
 \end{tabular}} 
 \end{ruledtabular}
\end{table}
\begin{table}[htbp]
 \centering
 \caption{List of possible $\mathcal{PT}$-symmetric mixed-parity pairing states in $D_{2h}$ point group. For each pairing state, IR of the order parameter, corresponding multipole moment, type of asymmetric modulation in the BS, and possible form of center-of-mass momentum of Cooper pairs $\bm{q}$ are shown. 
 IRs with odd time-reversal parity are denoted by $\Gamma^{-}$. 
 The anapole moment along the $\nu$-axis is expressed as $T_{\nu}$. On the other hand, $M_{0}$, $M_2$, $M_4$, ... denote nonpolar magnetic multipole moment, namely, the magnetic monopole, magnetic quadrupole, magnetic hexadecapole, ..., respectively.}
 \label{tab:UTe2}
 \begin{ruledtabular}
{\renewcommand \arraystretch{1.3}
 \begin{tabular}{ccccc}
  Pairing state & IR & Multipole & Modulation in BS& $\bm{q}$ of FFLO states \\ \hline
  $A_{g}+iA_{u}$, $B_{1g}+iB_{1u}$, $B_{2g}+iB_{2u}$, $B_{3g}+iB_{3u}$ & $A_{u}^{-}$ & $M_0$, $M_2$, $M_4$, ... & $k_xk_yk_z$ & $\bm{q}=(0,0,0)$ \\ 
    $A_{g}+iB_{1u}$, $B_{1g}+iA_{u}$, $B_{2g}+iB_{3u}$, $B_{3g}+iB_{2u}$ & $B_{1u}^{-}$ & $T_z$ &  $k_z$ & $\bm{q}=(0,0,q)$\\ 
    $A_{g}+iB_{2u}$, $B_{1g}+iB_{3u}$, $B_{2g}+iA_{u}$, $B_{3g}+iB_{1u}$ & $B_{2u}^{-}$ & $T_y$ &  $k_y$ & $\bm{q}=(0,q,0)$ \\ 
    $A_{g}+iB_{3u}$, $B_{1g}+iB_{2u}$, $B_{2g}+iB_{1u}$, $B_{3g}+iA_{u}$ & $B_{3u}^{-}$ & $T_x$ &  $k_x$ & $\bm{q}=(q,0,0)$ 
 \end{tabular}} 
 \end{ruledtabular}
\end{table}


\begin{thebibliography}{69}%
\makeatletter
\providecommand \@ifxundefined [1]{%
 \@ifx{#1\undefined}
}%
\providecommand \@ifnum [1]{%
 \ifnum #1\expandafter \@firstoftwo
 \else \expandafter \@secondoftwo
 \fi
}%
\providecommand \@ifx [1]{%
 \ifx #1\expandafter \@firstoftwo
 \else \expandafter \@secondoftwo
 \fi
}%
\providecommand \natexlab [1]{#1}%
\providecommand \enquote  [1]{``#1''}%
\providecommand \bibnamefont  [1]{#1}%
\providecommand \bibfnamefont [1]{#1}%
\providecommand \citenamefont [1]{#1}%
\providecommand \href@noop [0]{\@secondoftwo}%
\providecommand \href [0]{\begingroup \@sanitize@url \@href}%
\providecommand \@href[1]{\@@startlink{#1}\@@href}%
\providecommand \@@href[1]{\endgroup#1\@@endlink}%
\providecommand \@sanitize@url [0]{\catcode `\\12\catcode `\$12\catcode
  `\&12\catcode `\#12\catcode `\^12\catcode `\_12\catcode `\%12\relax}%
\providecommand \@@startlink[1]{}%
\providecommand \@@endlink[0]{}%
\providecommand \url  [0]{\begingroup\@sanitize@url \@url }%
\providecommand \@url [1]{\endgroup\@href {#1}{\urlprefix }}%
\providecommand \urlprefix  [0]{URL }%
\providecommand \Eprint [0]{\href }%
\providecommand \doibase [0]{http://dx.doi.org/}%
\providecommand \selectlanguage [0]{\@gobble}%
\providecommand \bibinfo  [0]{\@secondoftwo}%
\providecommand \bibfield  [0]{\@secondoftwo}%
\providecommand \translation [1]{[#1]}%
\providecommand \BibitemOpen [0]{}%
\providecommand \bibitemStop [0]{}%
\providecommand \bibitemNoStop [0]{.\EOS\space}%
\providecommand \EOS [0]{\spacefactor3000\relax}%
\providecommand \BibitemShut  [1]{\csname bibitem#1\endcsname}%
\let\auto@bib@innerbib\@empty
\bibitem [{\citenamefont {Sigrist}\ and\ \citenamefont
  {Ueda}(1991)}]{Sigrist-Ueda}%
  \BibitemOpen
  \bibfield  {author} {\bibinfo {author} {\bibfnamefont {M.}~\bibnamefont
  {Sigrist}}\ and\ \bibinfo {author} {\bibfnamefont {K.}~\bibnamefont {Ueda}},\
  }\href {\doibase 10.1103/RevModPhys.63.239} {\bibfield  {journal} {\bibinfo
  {journal} {Rev. Mod. Phys.}\ }\textbf {\bibinfo {volume} {63}},\ \bibinfo
  {pages} {239} (\bibinfo {year} {1991})}\BibitemShut {NoStop}%
\bibitem [{\citenamefont {Leggett}(1975)}]{Leggett1975}%
  \BibitemOpen
  \bibfield  {author} {\bibinfo {author} {\bibfnamefont {A.~J.}\ \bibnamefont
  {Leggett}},\ }\href {\doibase 10.1103/RevModPhys.47.331} {\bibfield
  {journal} {\bibinfo  {journal} {Rev. Mod. Phys.}\ }\textbf {\bibinfo {volume}
  {47}},\ \bibinfo {pages} {331} (\bibinfo {year} {1975})}\BibitemShut
  {NoStop}%
\bibitem [{\citenamefont {Hasan}\ and\ \citenamefont
  {Kane}(2010)}]{Hasan2010-topo}%
  \BibitemOpen
  \bibfield  {author} {\bibinfo {author} {\bibfnamefont {M.~Z.}\ \bibnamefont
  {Hasan}}\ and\ \bibinfo {author} {\bibfnamefont {C.~L.}\ \bibnamefont
  {Kane}},\ }\href {\doibase 10.1103/RevModPhys.82.3045} {\bibfield  {journal}
  {\bibinfo  {journal} {Rev. Mod. Phys.}\ }\textbf {\bibinfo {volume} {82}},\
  \bibinfo {pages} {3045} (\bibinfo {year} {2010})}\BibitemShut {NoStop}%
\bibitem [{\citenamefont {Qi}\ and\ \citenamefont {Zhang}(2011)}]{Qi2011-topo}%
  \BibitemOpen
  \bibfield  {author} {\bibinfo {author} {\bibfnamefont {X.-g.}\ \bibnamefont
  {Qi}}\ and\ \bibinfo {author} {\bibfnamefont {S.-C.}\ \bibnamefont {Zhang}},\
  }\href {\doibase 10.1103/RevModPhys.83.1057} {\bibfield  {journal} {\bibinfo
  {journal} {Rev. Mod. Phys.}\ }\textbf {\bibinfo {volume} {83}},\ \bibinfo
  {pages} {1057} (\bibinfo {year} {2011})}\BibitemShut {NoStop}%
\bibitem [{\citenamefont {Taylor}\ and\ \citenamefont
  {Kallin}(2012)}]{Taylor2012AHE}%
  \BibitemOpen
  \bibfield  {author} {\bibinfo {author} {\bibfnamefont {E.}~\bibnamefont
  {Taylor}}\ and\ \bibinfo {author} {\bibfnamefont {C.}~\bibnamefont
  {Kallin}},\ }\href {\doibase 10.1103/PhysRevLett.108.157001} {\bibfield
  {journal} {\bibinfo  {journal} {Phys. Rev. Lett.}\ }\textbf {\bibinfo
  {volume} {108}},\ \bibinfo {pages} {157001} (\bibinfo {year}
  {2012})}\BibitemShut {NoStop}%
\bibitem [{\citenamefont {Bauer}\ and\ \citenamefont
  {Sigrist}(2012)}]{Bauer2012NCS}%
  \BibitemOpen
  \bibinfo {editor} {\bibfnamefont {E.}~\bibnamefont {Bauer}}\ and\ \bibinfo
  {editor} {\bibfnamefont {M.}~\bibnamefont {Sigrist}},\ eds.,\ \href@noop {}
  {\emph {\bibinfo {title} {{Noncentrosymmetric Superconductor: Introduction
  and Overview}}}}\ (\bibinfo  {publisher} {Springer},\ \bibinfo {address}
  {Berlin},\ \bibinfo {year} {2012})\BibitemShut {NoStop}%
\bibitem [{\citenamefont {Smidman}\ \emph {et~al.}(2017)\citenamefont
  {Smidman}, \citenamefont {Salamon}, \citenamefont {Yuan},\ and\ \citenamefont
  {Agterberg}}]{Smidman2017}%
  \BibitemOpen
  \bibfield  {author} {\bibinfo {author} {\bibfnamefont {M.}~\bibnamefont
  {Smidman}}, \bibinfo {author} {\bibfnamefont {M.~B.}\ \bibnamefont
  {Salamon}}, \bibinfo {author} {\bibfnamefont {H.~Q.}\ \bibnamefont {Yuan}}, \
  and\ \bibinfo {author} {\bibfnamefont {D.~F.}\ \bibnamefont {Agterberg}},\
  }\href {\doibase 10.1088/1361-6633/80/3/036501} {\bibfield  {journal}
  {\bibinfo  {journal} {Rep. Prog. Phys.}\ }\textbf {\bibinfo {volume} {80}},\
  \bibinfo {pages} {036501} (\bibinfo {year} {2017})}\BibitemShut {NoStop}%
\bibitem [{\citenamefont {Wu}\ and\ \citenamefont
  {Hirsch}(2010)}]{ultracold2010}%
  \BibitemOpen
  \bibfield  {author} {\bibinfo {author} {\bibfnamefont {C.}~\bibnamefont
  {Wu}}\ and\ \bibinfo {author} {\bibfnamefont {J.~E.}\ \bibnamefont
  {Hirsch}},\ }\href {\doibase 10.1103/PhysRevB.81.020508} {\bibfield
  {journal} {\bibinfo  {journal} {Phys. Rev. B}\ }\textbf {\bibinfo {volume}
  {81}},\ \bibinfo {pages} {020508(R)} (\bibinfo {year} {2010})}\BibitemShut
  {NoStop}%
\bibitem [{\citenamefont {Zhou}\ \emph {et~al.}(2017)\citenamefont {Zhou},
  \citenamefont {Yi},\ and\ \citenamefont {Cui}}]{Zhou2017}%
  \BibitemOpen
  \bibfield  {author} {\bibinfo {author} {\bibfnamefont {L.~H.}\ \bibnamefont
  {Zhou}}, \bibinfo {author} {\bibfnamefont {W.}~\bibnamefont {Yi}}, \ and\
  \bibinfo {author} {\bibfnamefont {X.~L.}\ \bibnamefont {Cui}},\ }\href
  {\doibase 10.1007/s11433-017-9087-7} {\bibfield  {journal} {\bibinfo
  {journal} {Sci. China Phys. Mech. Astron.}\ }\textbf {\bibinfo {volume}
  {60}},\ \bibinfo {pages} {127011} (\bibinfo {year} {2017})}\BibitemShut
  {NoStop}%
\bibitem [{\citenamefont {Fu}(2015)}]{Fu2015_SOC}%
  \BibitemOpen
  \bibfield  {author} {\bibinfo {author} {\bibfnamefont {L.}~\bibnamefont
  {Fu}},\ }\href {\doibase 10.1103/PhysRevLett.115.026401} {\bibfield
  {journal} {\bibinfo  {journal} {Phys. Rev. Lett.}\ }\textbf {\bibinfo
  {volume} {115}},\ \bibinfo {pages} {026401} (\bibinfo {year}
  {2015})}\BibitemShut {NoStop}%
\bibitem [{\citenamefont {Kozii}\ and\ \citenamefont
  {Fu}(2015)}]{Kozii-Fu2015}%
  \BibitemOpen
  \bibfield  {author} {\bibinfo {author} {\bibfnamefont {V.}~\bibnamefont
  {Kozii}}\ and\ \bibinfo {author} {\bibfnamefont {L.}~\bibnamefont {Fu}},\
  }\href {\doibase 10.1103/PhysRevLett.115.207002} {\bibfield  {journal}
  {\bibinfo  {journal} {Phys. Rev. Lett.}\ }\textbf {\bibinfo {volume} {115}},\
  \bibinfo {pages} {207002} (\bibinfo {year} {2015})}\BibitemShut {NoStop}%
\bibitem [{\citenamefont {Sumita}\ and\ \citenamefont
  {Yanase}(2020)}]{Sumita2020}%
  \BibitemOpen
  \bibfield  {author} {\bibinfo {author} {\bibfnamefont {S.}~\bibnamefont
  {Sumita}}\ and\ \bibinfo {author} {\bibfnamefont {Y.}~\bibnamefont
  {Yanase}},\ }\href {\doibase 10.1103/PhysRevResearch.2.033225} {\bibfield
  {journal} {\bibinfo  {journal} {Phys. Rev. Research}\ }\textbf {\bibinfo
  {volume} {2}},\ \bibinfo {pages} {033225} (\bibinfo {year}
  {2020})}\BibitemShut {NoStop}%
\bibitem [{\citenamefont {Hiroi}\ \emph {et~al.}(2018)\citenamefont {Hiroi},
  \citenamefont {Yamaura}, \citenamefont {Kobayashi}, \citenamefont
  {Matsubayashi},\ and\ \citenamefont {Hirai}}]{Hiroi2018pyrochlore}%
  \BibitemOpen
  \bibfield  {author} {\bibinfo {author} {\bibfnamefont {Z.}~\bibnamefont
  {Hiroi}}, \bibinfo {author} {\bibfnamefont {J.}~\bibnamefont {Yamaura}},
  \bibinfo {author} {\bibfnamefont {T.~C.}\ \bibnamefont {Kobayashi}}, \bibinfo
  {author} {\bibfnamefont {Y.}~\bibnamefont {Matsubayashi}}, \ and\ \bibinfo
  {author} {\bibfnamefont {D.}~\bibnamefont {Hirai}},\ }\href
  {https://journals.jps.jp/doi/10.7566/JPSJ.87.024702} {\bibfield  {journal}
  {\bibinfo  {journal} {J. Phys. Soc. Jpn.}\ }\textbf {\bibinfo {volume}
  {87}},\ \bibinfo {pages} {024702} (\bibinfo {year} {2018})}\BibitemShut
  {NoStop}%
\bibitem [{\citenamefont {Schumann}\ \emph {et~al.}(2020)\citenamefont
  {Schumann}, \citenamefont {Galletti}, \citenamefont {Jeong}, \citenamefont
  {Ahadi}, \citenamefont {Strickland}, \citenamefont {Salmani-Rezaie},\ and\
  \citenamefont {Stemmer}}]{STO_parity}%
  \BibitemOpen
  \bibfield  {author} {\bibinfo {author} {\bibfnamefont {T.}~\bibnamefont
  {Schumann}}, \bibinfo {author} {\bibfnamefont {L.}~\bibnamefont {Galletti}},
  \bibinfo {author} {\bibfnamefont {H.}~\bibnamefont {Jeong}}, \bibinfo
  {author} {\bibfnamefont {K.}~\bibnamefont {Ahadi}}, \bibinfo {author}
  {\bibfnamefont {W.~M.}\ \bibnamefont {Strickland}}, \bibinfo {author}
  {\bibfnamefont {S.}~\bibnamefont {Salmani-Rezaie}}, \ and\ \bibinfo {author}
  {\bibfnamefont {S.}~\bibnamefont {Stemmer}},\ }\href {\doibase
  10.1103/PhysRevB.101.100503} {\bibfield  {journal} {\bibinfo  {journal}
  {Phys. Rev. B}\ }\textbf {\bibinfo {volume} {101}},\ \bibinfo {pages}
  {100503(R)} (\bibinfo {year} {2020})}\BibitemShut {NoStop}%
\bibitem [{\citenamefont {Wang}\ and\ \citenamefont {Fu}(2017)}]{WangFu2017}%
  \BibitemOpen
  \bibfield  {author} {\bibinfo {author} {\bibfnamefont {Y.}~\bibnamefont
  {Wang}}\ and\ \bibinfo {author} {\bibfnamefont {L.}~\bibnamefont {Fu}},\
  }\href {\doibase 10.1103/PhysRevLett.119.187003} {\bibfield  {journal}
  {\bibinfo  {journal} {Phys. Rev. Lett.}\ }\textbf {\bibinfo {volume} {119}},\
  \bibinfo {pages} {187003} (\bibinfo {year} {2017})}\BibitemShut {NoStop}%
\bibitem [{\citenamefont {Yang}\ \emph {et~al.}(2020)\citenamefont {Yang},
  \citenamefont {Xu},\ and\ \citenamefont {Wu}}]{Wang2020}%
  \BibitemOpen
  \bibfield  {author} {\bibinfo {author} {\bibfnamefont {W.}~\bibnamefont
  {Yang}}, \bibinfo {author} {\bibfnamefont {C.}~\bibnamefont {Xu}}, \ and\
  \bibinfo {author} {\bibfnamefont {C.}~\bibnamefont {Wu}},\ }\href {\doibase
  10.1103/PhysRevResearch.2.042047} {\bibfield  {journal} {\bibinfo  {journal}
  {Phys. Rev. Research}\ }\textbf {\bibinfo {volume} {2}},\ \bibinfo {pages}
  {042047(R)} (\bibinfo {year} {2020})}\BibitemShut {NoStop}%
\bibitem [{\citenamefont {Ryu}\ \emph {et~al.}(2012)\citenamefont {Ryu},
  \citenamefont {Moore},\ and\ \citenamefont {Ludwig}}]{Ryu2012}%
  \BibitemOpen
  \bibfield  {author} {\bibinfo {author} {\bibfnamefont {S.}~\bibnamefont
  {Ryu}}, \bibinfo {author} {\bibfnamefont {J.~E.}\ \bibnamefont {Moore}}, \
  and\ \bibinfo {author} {\bibfnamefont {A.~W.~W.}\ \bibnamefont {Ludwig}},\
  }\href {\doibase 10.1103/PhysRevB.85.045104} {\bibfield  {journal} {\bibinfo
  {journal} {Phys. Rev. B}\ }\textbf {\bibinfo {volume} {85}},\ \bibinfo
  {pages} {045104} (\bibinfo {year} {2012})}\BibitemShut {NoStop}%
\bibitem [{\citenamefont {Qi}\ \emph {et~al.}(2013)\citenamefont {Qi},
  \citenamefont {Witten},\ and\ \citenamefont {Zhang}}]{Qi2013}%
  \BibitemOpen
  \bibfield  {author} {\bibinfo {author} {\bibfnamefont {X.-g.}\ \bibnamefont
  {Qi}}, \bibinfo {author} {\bibfnamefont {E.}~\bibnamefont {Witten}}, \ and\
  \bibinfo {author} {\bibfnamefont {S.-C.}\ \bibnamefont {Zhang}},\ }\href
  {\doibase 10.1103/PhysRevB.87.134519} {\bibfield  {journal} {\bibinfo
  {journal} {Phys. Rev. B}\ }\textbf {\bibinfo {volume} {87}},\ \bibinfo
  {pages} {134519} (\bibinfo {year} {2013})}\BibitemShut {NoStop}%
\bibitem [{\citenamefont {Goswami}\ and\ \citenamefont
  {Roy}(2014)}]{Goswami2014}%
  \BibitemOpen
  \bibfield  {author} {\bibinfo {author} {\bibfnamefont {P.}~\bibnamefont
  {Goswami}}\ and\ \bibinfo {author} {\bibfnamefont {B.}~\bibnamefont {Roy}},\
  }\href {\doibase 10.1103/PhysRevB.90.041301} {\bibfield  {journal} {\bibinfo
  {journal} {Phys. Rev. B}\ }\textbf {\bibinfo {volume} {90}},\ \bibinfo
  {pages} {041301(R)} (\bibinfo {year} {2014})}\BibitemShut {NoStop}%
\bibitem [{\citenamefont {Shiozaki}\ and\ \citenamefont
  {Fujimoto}(2014)}]{Shiozaki2014}%
  \BibitemOpen
  \bibfield  {author} {\bibinfo {author} {\bibfnamefont {K.}~\bibnamefont
  {Shiozaki}}\ and\ \bibinfo {author} {\bibfnamefont {S.}~\bibnamefont
  {Fujimoto}},\ }\href {\doibase 10.1103/PhysRevB.89.054506} {\bibfield
  {journal} {\bibinfo  {journal} {Phys. Rev. B}\ }\textbf {\bibinfo {volume}
  {89}},\ \bibinfo {pages} {054506} (\bibinfo {year} {2014})}\BibitemShut
  {NoStop}%
\bibitem [{\citenamefont {Roy}(2020)}]{Roy2020}%
  \BibitemOpen
  \bibfield  {author} {\bibinfo {author} {\bibfnamefont {B.}~\bibnamefont
  {Roy}},\ }\href {\doibase 10.1103/PhysRevB.101.220506} {\bibfield  {journal}
  {\bibinfo  {journal} {Phys. Rev. B}\ }\textbf {\bibinfo {volume} {101}},\
  \bibinfo {pages} {220506(R)} (\bibinfo {year} {2020})}\BibitemShut {NoStop}%
\bibitem [{\citenamefont {Xu}\ and\ \citenamefont {Yang}()}]{Xu2020axion}%
  \BibitemOpen
  \bibfield  {author} {\bibinfo {author} {\bibfnamefont {C.}~\bibnamefont
  {Xu}}\ and\ \bibinfo {author} {\bibfnamefont {W.}~\bibnamefont {Yang}},\
  }\href {https://arxiv.org/abs/2009.12998} {\ }\Eprint
  {http://arxiv.org/abs/2009.12998} {arXiv:2009.12998} \BibitemShut {NoStop}%
\bibitem [{\citenamefont {Scaffidi}()}]{SRO_mixed-parity}%
  \BibitemOpen
  \bibfield  {author} {\bibinfo {author} {\bibfnamefont {T.}~\bibnamefont
  {Scaffidi}},\ }\href {https://arxiv.org/abs/2007.13769} {\ }\Eprint
  {http://arxiv.org/abs/2007.13769} {arXiv:2007.13769} \BibitemShut {NoStop}%
\bibitem [{\citenamefont {Ishizuka}\ and\ \citenamefont
  {Yanase}(2021)}]{Ishizuka2020}%
  \BibitemOpen
  \bibfield  {author} {\bibinfo {author} {\bibfnamefont {J.}~\bibnamefont
  {Ishizuka}}\ and\ \bibinfo {author} {\bibfnamefont {Y.}~\bibnamefont
  {Yanase}},\ }\href {\doibase 10.1103/PhysRevB.103.094504} {\bibfield
  {journal} {\bibinfo  {journal} {Phys. Rev. B}\ }\textbf {\bibinfo {volume}
  {103}},\ \bibinfo {pages} {094504} (\bibinfo {year} {2021})}\BibitemShut
  {NoStop}%
\bibitem [{\citenamefont {Braithwaite}\ \emph {et~al.}(2019)\citenamefont
  {Braithwaite}, \citenamefont {Vali{\v{s}}ka}, \citenamefont {Knebel},
  \citenamefont {Lapertot}, \citenamefont {Brison}, \citenamefont {Pourret},
  \citenamefont {Zhitomirsky}, \citenamefont {Flouquet}, \citenamefont
  {Honda},\ and\ \citenamefont {Aoki}}]{Braithwaite2019}%
  \BibitemOpen
  \bibfield  {author} {\bibinfo {author} {\bibfnamefont {D.}~\bibnamefont
  {Braithwaite}}, \bibinfo {author} {\bibfnamefont {M.}~\bibnamefont
  {Vali{\v{s}}ka}}, \bibinfo {author} {\bibfnamefont {G.}~\bibnamefont
  {Knebel}}, \bibinfo {author} {\bibfnamefont {G.}~\bibnamefont {Lapertot}},
  \bibinfo {author} {\bibfnamefont {J.~P.}\ \bibnamefont {Brison}}, \bibinfo
  {author} {\bibfnamefont {A.}~\bibnamefont {Pourret}}, \bibinfo {author}
  {\bibfnamefont {M.~E.}\ \bibnamefont {Zhitomirsky}}, \bibinfo {author}
  {\bibfnamefont {J.}~\bibnamefont {Flouquet}}, \bibinfo {author}
  {\bibfnamefont {F.}~\bibnamefont {Honda}}, \ and\ \bibinfo {author}
  {\bibfnamefont {D.}~\bibnamefont {Aoki}},\ }\href {\doibase
  10.1038/s42005-019-0248-z} {\bibfield  {journal} {\bibinfo  {journal}
  {Communications Physics}\ }\textbf {\bibinfo {volume} {2}},\ \bibinfo {pages}
  {1} (\bibinfo {year} {2019})}\BibitemShut {NoStop}%
\bibitem [{\citenamefont {Lin}\ \emph {et~al.}(2020)\citenamefont {Lin},
  \citenamefont {Campbell}, \citenamefont {Ran}, \citenamefont {Liu},
  \citenamefont {Kim}, \citenamefont {Nevidomskyy}, \citenamefont {Graf},
  \citenamefont {Butch},\ and\ \citenamefont {Paglione}}]{Lin2020}%
  \BibitemOpen
  \bibfield  {author} {\bibinfo {author} {\bibfnamefont {W.~C.}\ \bibnamefont
  {Lin}}, \bibinfo {author} {\bibfnamefont {D.~J.}\ \bibnamefont {Campbell}},
  \bibinfo {author} {\bibfnamefont {S.}~\bibnamefont {Ran}}, \bibinfo {author}
  {\bibfnamefont {I.~L.}\ \bibnamefont {Liu}}, \bibinfo {author} {\bibfnamefont
  {H.}~\bibnamefont {Kim}}, \bibinfo {author} {\bibfnamefont {A.~H.}\
  \bibnamefont {Nevidomskyy}}, \bibinfo {author} {\bibfnamefont
  {D.}~\bibnamefont {Graf}}, \bibinfo {author} {\bibfnamefont {N.~P.}\
  \bibnamefont {Butch}}, \ and\ \bibinfo {author} {\bibfnamefont
  {J.}~\bibnamefont {Paglione}},\ }\href {\doibase 10.1038/s41535-020-00270-w}
  {\bibfield  {journal} {\bibinfo  {journal} {npj Quantum Materials}\ }\textbf
  {\bibinfo {volume} {5}},\ \bibinfo {pages} {1} (\bibinfo {year}
  {2020})}\BibitemShut {NoStop}%
\bibitem [{\citenamefont {Aoki}\ \emph {et~al.}(2020)\citenamefont {Aoki},
  \citenamefont {Honda}, \citenamefont {Knebel}, \citenamefont {Braithwaite},
  \citenamefont {Nakamura}, \citenamefont {Li}, \citenamefont {Homma},
  \citenamefont {Shimizu}, \citenamefont {Sato}, \citenamefont {Brison},\ and\
  \citenamefont {Flouquet}}]{Aoki2020}%
  \BibitemOpen
  \bibfield  {author} {\bibinfo {author} {\bibfnamefont {D.}~\bibnamefont
  {Aoki}}, \bibinfo {author} {\bibfnamefont {F.}~\bibnamefont {Honda}},
  \bibinfo {author} {\bibfnamefont {G.}~\bibnamefont {Knebel}}, \bibinfo
  {author} {\bibfnamefont {D.}~\bibnamefont {Braithwaite}}, \bibinfo {author}
  {\bibfnamefont {A.}~\bibnamefont {Nakamura}}, \bibinfo {author}
  {\bibfnamefont {D.}~\bibnamefont {Li}}, \bibinfo {author} {\bibfnamefont
  {Y.}~\bibnamefont {Homma}}, \bibinfo {author} {\bibfnamefont
  {Y.}~\bibnamefont {Shimizu}}, \bibinfo {author} {\bibfnamefont {Y.~J.}\
  \bibnamefont {Sato}}, \bibinfo {author} {\bibfnamefont {J.-P.}\ \bibnamefont
  {Brison}}, \ and\ \bibinfo {author} {\bibfnamefont {J.}~\bibnamefont
  {Flouquet}},\ }\href {\doibase 10.7566/JPSJ.89.053705} {\bibfield  {journal}
  {\bibinfo  {journal} {J. Phys. Soc. Jpn.}\ }\textbf {\bibinfo {volume}
  {89}},\ \bibinfo {pages} {053705} (\bibinfo {year} {2020})}\BibitemShut
  {NoStop}%
\bibitem [{\citenamefont {Aoki}\ \emph {et~al.}(2021)\citenamefont {Aoki},
  \citenamefont {Kimata}, \citenamefont {Sato}, \citenamefont {Knebel},
  \citenamefont {Honda}, \citenamefont {Nakamura}, \citenamefont {Li},
  \citenamefont {Homma}, \citenamefont {Shimizu}, \citenamefont {Knafo},
  \citenamefont {Braithwaite}, \citenamefont {Vali邸ka}, \citenamefont
  {Pourret}, \citenamefont {Brison},\ and\ \citenamefont
  {Flouquet}}]{aoki2021fieldinduced}%
  \BibitemOpen
  \bibfield  {author} {\bibinfo {author} {\bibfnamefont {D.}~\bibnamefont
  {Aoki}}, \bibinfo {author} {\bibfnamefont {M.}~\bibnamefont {Kimata}},
  \bibinfo {author} {\bibfnamefont {Y.~J.}\ \bibnamefont {Sato}}, \bibinfo
  {author} {\bibfnamefont {G.}~\bibnamefont {Knebel}}, \bibinfo {author}
  {\bibfnamefont {F.}~\bibnamefont {Honda}}, \bibinfo {author} {\bibfnamefont
  {A.}~\bibnamefont {Nakamura}}, \bibinfo {author} {\bibfnamefont
  {D.}~\bibnamefont {Li}}, \bibinfo {author} {\bibfnamefont {Y.}~\bibnamefont
  {Homma}}, \bibinfo {author} {\bibfnamefont {Y.}~\bibnamefont {Shimizu}},
  \bibinfo {author} {\bibfnamefont {W.}~\bibnamefont {Knafo}}, \bibinfo
  {author} {\bibfnamefont {D.}~\bibnamefont {Braithwaite}}, \bibinfo {author}
  {\bibfnamefont {M.}~\bibnamefont {Vali{\v{s}}ka}}, \bibinfo {author}
  {\bibfnamefont {A.}~\bibnamefont {Pourret}}, \bibinfo {author} {\bibfnamefont
  {J.-P.}\ \bibnamefont {Brison}}, \ and\ \bibinfo {author} {\bibfnamefont
  {J.}~\bibnamefont {Flouquet}},\ }\href {\doibase 10.7566/JPSJ.90.074705}
  {\bibfield  {journal} {\bibinfo  {journal} {Journal of the Physical Society
  of Japan}\ }\textbf {\bibinfo {volume} {90}},\ \bibinfo {pages} {074705}
  (\bibinfo {year} {2021})}\BibitemShut {NoStop}%
\bibitem [{\citenamefont {Black-Schaffer}\ and\ \citenamefont
  {Balatsky}(2013)}]{AMB_AVB_2013}%
  \BibitemOpen
  \bibfield  {author} {\bibinfo {author} {\bibfnamefont {A.~M.}\ \bibnamefont
  {Black-Schaffer}}\ and\ \bibinfo {author} {\bibfnamefont {A.~V.}\
  \bibnamefont {Balatsky}},\ }\href {\doibase 10.1103/PhysRevB.88.104514}
  {\bibfield  {journal} {\bibinfo  {journal} {Phys. Rev. B}\ }\textbf {\bibinfo
  {volume} {88}},\ \bibinfo {pages} {104514} (\bibinfo {year}
  {2013})}\BibitemShut {NoStop}%
\bibitem [{\citenamefont {Wang}\ \emph {et~al.}(2017)\citenamefont {Wang},
  \citenamefont {Berlinsky}, \citenamefont {Zwicknagl},\ and\ \citenamefont
  {Kallin}}]{Wang2017}%
  \BibitemOpen
  \bibfield  {author} {\bibinfo {author} {\bibfnamefont {Z.}~\bibnamefont
  {Wang}}, \bibinfo {author} {\bibfnamefont {J.}~\bibnamefont {Berlinsky}},
  \bibinfo {author} {\bibfnamefont {G.}~\bibnamefont {Zwicknagl}}, \ and\
  \bibinfo {author} {\bibfnamefont {C.}~\bibnamefont {Kallin}},\ }\href
  {\doibase 10.1103/PhysRevB.96.174511} {\bibfield  {journal} {\bibinfo
  {journal} {Phys. Rev. B}\ }\textbf {\bibinfo {volume} {96}},\ \bibinfo
  {pages} {174511} (\bibinfo {year} {2017})}\BibitemShut {NoStop}%
\bibitem [{\citenamefont {Brydon}\ \emph {et~al.}(2019)\citenamefont {Brydon},
  \citenamefont {Abergel}, \citenamefont {Agterberg},\ and\ \citenamefont
  {Yakovenko}}]{Brydon2019AHE}%
  \BibitemOpen
  \bibfield  {author} {\bibinfo {author} {\bibfnamefont {P.~M.~R.}\
  \bibnamefont {Brydon}}, \bibinfo {author} {\bibfnamefont {D.~S.~L.}\
  \bibnamefont {Abergel}}, \bibinfo {author} {\bibfnamefont {D.~F.}\
  \bibnamefont {Agterberg}}, \ and\ \bibinfo {author} {\bibfnamefont {V.~M.}\
  \bibnamefont {Yakovenko}},\ }\href {\doibase 10.1103/PhysRevX.9.031025}
  {\bibfield  {journal} {\bibinfo  {journal} {Phys. Rev. X}\ }\textbf {\bibinfo
  {volume} {9}},\ \bibinfo {pages} {031025} (\bibinfo {year}
  {2019})}\BibitemShut {NoStop}%
\bibitem [{\citenamefont {Denys}\ and\ \citenamefont
  {Brydon}(2021)}]{Brydon2021AHE}%
  \BibitemOpen
  \bibfield  {author} {\bibinfo {author} {\bibfnamefont {M.~D.~E.}\
  \bibnamefont {Denys}}\ and\ \bibinfo {author} {\bibfnamefont {P.~M.~R.}\
  \bibnamefont {Brydon}},\ }\href {\doibase 10.1103/PhysRevB.103.094503}
  {\bibfield  {journal} {\bibinfo  {journal} {Phys. Rev. B}\ }\textbf {\bibinfo
  {volume} {103}},\ \bibinfo {pages} {094503} (\bibinfo {year}
  {2021})}\BibitemShut {NoStop}%
\bibitem [{\citenamefont {Triola}\ and\ \citenamefont
  {Black-Schaffer}(2018)}]{Triola2018}%
  \BibitemOpen
  \bibfield  {author} {\bibinfo {author} {\bibfnamefont {C.}~\bibnamefont
  {Triola}}\ and\ \bibinfo {author} {\bibfnamefont {A.~M.}\ \bibnamefont
  {Black-Schaffer}},\ }\href {\doibase 10.1103/PhysRevB.97.064505} {\bibfield
  {journal} {\bibinfo  {journal} {Phys. Rev. B}\ }\textbf {\bibinfo {volume}
  {97}},\ \bibinfo {pages} {064505} (\bibinfo {year} {2018})}\BibitemShut
  {NoStop}%
\bibitem [{\citenamefont {Agterberg}\ \emph {et~al.}(2017)\citenamefont
  {Agterberg}, \citenamefont {Brydon},\ and\ \citenamefont
  {Timm}}]{Agterberg_BogoFS}%
  \BibitemOpen
  \bibfield  {author} {\bibinfo {author} {\bibfnamefont {D.~F.}\ \bibnamefont
  {Agterberg}}, \bibinfo {author} {\bibfnamefont {P.~M.~R.}\ \bibnamefont
  {Brydon}}, \ and\ \bibinfo {author} {\bibfnamefont {C.}~\bibnamefont
  {Timm}},\ }\href {\doibase 10.1103/PhysRevLett.118.127001} {\bibfield
  {journal} {\bibinfo  {journal} {Phys. Rev. Lett.}\ }\textbf {\bibinfo
  {volume} {118}},\ \bibinfo {pages} {127001} (\bibinfo {year}
  {2017})}\BibitemShut {NoStop}%
\bibitem [{\citenamefont {Brydon}\ \emph {et~al.}(2018)\citenamefont {Brydon},
  \citenamefont {Agterberg}, \citenamefont {Menke},\ and\ \citenamefont
  {Timm}}]{Brydon_BogoFS}%
  \BibitemOpen
  \bibfield  {author} {\bibinfo {author} {\bibfnamefont {P.~M.~R.}\
  \bibnamefont {Brydon}}, \bibinfo {author} {\bibfnamefont {D.~F.}\
  \bibnamefont {Agterberg}}, \bibinfo {author} {\bibfnamefont {H.}~\bibnamefont
  {Menke}}, \ and\ \bibinfo {author} {\bibfnamefont {C.}~\bibnamefont {Timm}},\
  }\href {\doibase 10.1103/PhysRevB.98.224509} {\bibfield  {journal} {\bibinfo
  {journal} {Phys. Rev. B}\ }\textbf {\bibinfo {volume} {98}},\ \bibinfo
  {pages} {224509} (\bibinfo {year} {2018})}\BibitemShut {NoStop}%
\bibitem [{\citenamefont {Fulde}\ and\ \citenamefont {Ferrell}(1964)}]{FF1964}%
  \BibitemOpen
  \bibfield  {author} {\bibinfo {author} {\bibfnamefont {P.}~\bibnamefont
  {Fulde}}\ and\ \bibinfo {author} {\bibfnamefont {R.~A.}\ \bibnamefont
  {Ferrell}},\ }\href {\doibase 10.1103/PhysRev.135.A550} {\bibfield  {journal}
  {\bibinfo  {journal} {Phys. Rev.}\ }\textbf {\bibinfo {volume} {135}},\
  \bibinfo {pages} {A550} (\bibinfo {year} {1964})}\BibitemShut {NoStop}%
\bibitem [{\citenamefont {Larkin}\ and\ \citenamefont
  {Ovchinnikov}(1964)}]{LO1964}%
  \BibitemOpen
  \bibfield  {author} {\bibinfo {author} {\bibfnamefont {A.~I.}\ \bibnamefont
  {Larkin}}\ and\ \bibinfo {author} {\bibfnamefont {Y.~N.}\ \bibnamefont
  {Ovchinnikov}},\ }\href@noop {} {\bibfield  {journal} {\bibinfo  {journal}
  {Zh. Eksp. Teor. Fiz.}\ }\textbf {\bibinfo {volume} {47}},\ \bibinfo {pages}
  {1136} (\bibinfo {year} {1964})},\ \bibinfo {note} {{[translation: Sov. Phys.
  JETP {\bfseries 20}, 762 (1965)]}}\BibitemShut {NoStop}%
\bibitem [{\citenamefont {Mineev}\ and\ \citenamefont
  {Samokhin}(2008)}]{Mineev_Lifshitz}%
  \BibitemOpen
  \bibfield  {author} {\bibinfo {author} {\bibfnamefont {V.~P.}\ \bibnamefont
  {Mineev}}\ and\ \bibinfo {author} {\bibfnamefont {K.~V.}\ \bibnamefont
  {Samokhin}},\ }\href {\doibase 10.1103/PhysRevB.78.144503} {\bibfield
  {journal} {\bibinfo  {journal} {Phys. Rev. B}\ }\textbf {\bibinfo {volume}
  {78}},\ \bibinfo {pages} {144503} (\bibinfo {year} {2008})}\BibitemShut
  {NoStop}%
\bibitem [{\citenamefont {Spaldin}\ \emph {et~al.}(2008)\citenamefont
  {Spaldin}, \citenamefont {Fiebig},\ and\ \citenamefont
  {Mostovoy}}]{Spaldin_2008}%
  \BibitemOpen
  \bibfield  {author} {\bibinfo {author} {\bibfnamefont {N.~A.}\ \bibnamefont
  {Spaldin}}, \bibinfo {author} {\bibfnamefont {M.}~\bibnamefont {Fiebig}}, \
  and\ \bibinfo {author} {\bibfnamefont {M.}~\bibnamefont {Mostovoy}},\ }\href
  {\doibase 10.1088/0953-8984/20/43/434203} {\bibfield  {journal} {\bibinfo
  {journal} {Journal of Physics: Condensed Matter}\ }\textbf {\bibinfo {volume}
  {20}},\ \bibinfo {pages} {434203} (\bibinfo {year} {2008})}\BibitemShut
  {NoStop}%
\bibitem [{\citenamefont {Flambaum}\ \emph {et~al.}(1984)\citenamefont
  {Flambaum}, \citenamefont {Khriplovich},\ and\ \citenamefont
  {Sushkov}}]{flambaum1984nuclear}%
  \BibitemOpen
  \bibfield  {author} {\bibinfo {author} {\bibfnamefont {V.~V.}\ \bibnamefont
  {Flambaum}}, \bibinfo {author} {\bibfnamefont {I.~B.}\ \bibnamefont
  {Khriplovich}}, \ and\ \bibinfo {author} {\bibfnamefont {O.~P.}\ \bibnamefont
  {Sushkov}},\ }\href
  {https://www.sciencedirect.com/science/article/abs/pii/0370269384901400}
  {\bibfield  {journal} {\bibinfo  {journal} {Physics Letters B}\ }\textbf
  {\bibinfo {volume} {146}},\ \bibinfo {pages} {367} (\bibinfo {year}
  {1984})}\BibitemShut {NoStop}%
\bibitem [{\citenamefont {Jeong}\ \emph {et~al.}(2017)\citenamefont {Jeong},
  \citenamefont {Sidis}, \citenamefont {Louat}, \citenamefont {Brouet},\ and\
  \citenamefont {Bourges}}]{Jeong2017}%
  \BibitemOpen
  \bibfield  {author} {\bibinfo {author} {\bibfnamefont {J.}~\bibnamefont
  {Jeong}}, \bibinfo {author} {\bibfnamefont {Y.}~\bibnamefont {Sidis}},
  \bibinfo {author} {\bibfnamefont {A.}~\bibnamefont {Louat}}, \bibinfo
  {author} {\bibfnamefont {V.}~\bibnamefont {Brouet}}, \ and\ \bibinfo {author}
  {\bibfnamefont {P.}~\bibnamefont {Bourges}},\ }\href {\doibase
  10.1038/ncomms15119} {\bibfield  {journal} {\bibinfo  {journal} {Nature
  Communications}\ }\textbf {\bibinfo {volume} {8}},\ \bibinfo {pages} {15119}
  (\bibinfo {year} {2017})}\BibitemShut {NoStop}%
\bibitem [{\citenamefont {Murayama}\ \emph {et~al.}(2021)\citenamefont
  {Murayama}, \citenamefont {Ishida}, \citenamefont {Kurihara}, \citenamefont
  {Ono}, \citenamefont {Sato}, \citenamefont {Kasahara}, \citenamefont
  {Watanabe}, \citenamefont {Yanase}, \citenamefont {Cao}, \citenamefont
  {Mizukami}, \citenamefont {Shibauchi}, \citenamefont {Matsuda},\ and\
  \citenamefont {Kasahara}}]{Murayama2021}%
  \BibitemOpen
  \bibfield  {author} {\bibinfo {author} {\bibfnamefont {H.}~\bibnamefont
  {Murayama}}, \bibinfo {author} {\bibfnamefont {K.}~\bibnamefont {Ishida}},
  \bibinfo {author} {\bibfnamefont {R.}~\bibnamefont {Kurihara}}, \bibinfo
  {author} {\bibfnamefont {T.}~\bibnamefont {Ono}}, \bibinfo {author}
  {\bibfnamefont {Y.}~\bibnamefont {Sato}}, \bibinfo {author} {\bibfnamefont
  {Y.}~\bibnamefont {Kasahara}}, \bibinfo {author} {\bibfnamefont
  {H.}~\bibnamefont {Watanabe}}, \bibinfo {author} {\bibfnamefont
  {Y.}~\bibnamefont {Yanase}}, \bibinfo {author} {\bibfnamefont
  {G.}~\bibnamefont {Cao}}, \bibinfo {author} {\bibfnamefont {Y.}~\bibnamefont
  {Mizukami}}, \bibinfo {author} {\bibfnamefont {T.}~\bibnamefont {Shibauchi}},
  \bibinfo {author} {\bibfnamefont {Y.}~\bibnamefont {Matsuda}}, \ and\
  \bibinfo {author} {\bibfnamefont {S.}~\bibnamefont {Kasahara}},\ }\href
  {\doibase 10.1103/PhysRevX.11.011021} {\bibfield  {journal} {\bibinfo
  {journal} {Phys. Rev. X}\ }\textbf {\bibinfo {volume} {11}},\ \bibinfo
  {pages} {011021} (\bibinfo {year} {2021})}\BibitemShut {NoStop}%
\bibitem [{\citenamefont {Watanabe}\ and\ \citenamefont
  {Yanase}(2021)}]{HW2021_Photocurrent}%
  \BibitemOpen
  \bibfield  {author} {\bibinfo {author} {\bibfnamefont {H.}~\bibnamefont
  {Watanabe}}\ and\ \bibinfo {author} {\bibfnamefont {Y.}~\bibnamefont
  {Yanase}},\ }\href {\doibase 10.1103/PhysRevX.11.011001} {\bibfield
  {journal} {\bibinfo  {journal} {Phys. Rev. X}\ }\textbf {\bibinfo {volume}
  {11}},\ \bibinfo {pages} {011001} (\bibinfo {year} {2021})}\BibitemShut
  {NoStop}%
\bibitem [{\citenamefont {Ahn}\ \emph {et~al.}(2020)\citenamefont {Ahn},
  \citenamefont {Guo},\ and\ \citenamefont {Nagaosa}}]{Ahn2020}%
  \BibitemOpen
  \bibfield  {author} {\bibinfo {author} {\bibfnamefont {J.}~\bibnamefont
  {Ahn}}, \bibinfo {author} {\bibfnamefont {G.-Y.}\ \bibnamefont {Guo}}, \ and\
  \bibinfo {author} {\bibfnamefont {N.}~\bibnamefont {Nagaosa}},\ }\href
  {\doibase 10.1103/PhysRevX.10.041041} {\bibfield  {journal} {\bibinfo
  {journal} {Phys. Rev. X}\ }\textbf {\bibinfo {volume} {10}},\ \bibinfo
  {pages} {041041} (\bibinfo {year} {2020})}\BibitemShut {NoStop}%
\bibitem [{\citenamefont {Agterberg}\ and\ \citenamefont
  {Kaur}(2007)}]{Agterberg2007FFLO}%
  \BibitemOpen
  \bibfield  {author} {\bibinfo {author} {\bibfnamefont {D.~F.}\ \bibnamefont
  {Agterberg}}\ and\ \bibinfo {author} {\bibfnamefont {R.~P.}\ \bibnamefont
  {Kaur}},\ }\href {\doibase 10.1103/PhysRevB.75.064511} {\bibfield  {journal}
  {\bibinfo  {journal} {Phys. Rev. B}\ }\textbf {\bibinfo {volume} {75}},\
  \bibinfo {pages} {064511} (\bibinfo {year} {2007})}\BibitemShut {NoStop}%
\bibitem [{\citenamefont {Sumita}\ and\ \citenamefont
  {Yanase}(2016)}]{Sumita2016}%
  \BibitemOpen
  \bibfield  {author} {\bibinfo {author} {\bibfnamefont {S.}~\bibnamefont
  {Sumita}}\ and\ \bibinfo {author} {\bibfnamefont {Y.}~\bibnamefont
  {Yanase}},\ }\href {\doibase 10.1103/PhysRevB.93.224507} {\bibfield
  {journal} {\bibinfo  {journal} {Phys. Rev. B}\ }\textbf {\bibinfo {volume}
  {93}},\ \bibinfo {pages} {224507} (\bibinfo {year} {2016})}\BibitemShut
  {NoStop}%
\bibitem [{\citenamefont {Sumita}\ \emph {et~al.}(2017)\citenamefont {Sumita},
  \citenamefont {Nomoto},\ and\ \citenamefont {Yanase}}]{Sumita2017}%
  \BibitemOpen
  \bibfield  {author} {\bibinfo {author} {\bibfnamefont {S.}~\bibnamefont
  {Sumita}}, \bibinfo {author} {\bibfnamefont {T.}~\bibnamefont {Nomoto}}, \
  and\ \bibinfo {author} {\bibfnamefont {Y.}~\bibnamefont {Yanase}},\ }\href
  {\doibase 10.1103/PhysRevLett.119.027001} {\bibfield  {journal} {\bibinfo
  {journal} {Phys. Rev. Lett.}\ }\textbf {\bibinfo {volume} {119}},\ \bibinfo
  {pages} {027001} (\bibinfo {year} {2017})}\BibitemShut {NoStop}%
\bibitem [{\citenamefont {Timm}\ \emph {et~al.}(2021)\citenamefont {Timm},
  \citenamefont {Brydon},\ and\ \citenamefont
  {Agterberg}}]{Timm2021Distortional}%
  \BibitemOpen
  \bibfield  {author} {\bibinfo {author} {\bibfnamefont {C.}~\bibnamefont
  {Timm}}, \bibinfo {author} {\bibfnamefont {P.~M.~R.}\ \bibnamefont {Brydon}},
  \ and\ \bibinfo {author} {\bibfnamefont {D.~F.}\ \bibnamefont {Agterberg}},\
  }\href {\doibase 10.1103/PhysRevB.103.024521} {\bibfield  {journal} {\bibinfo
   {journal} {Phys. Rev. B}\ }\textbf {\bibinfo {volume} {103}},\ \bibinfo
  {pages} {024521} (\bibinfo {year} {2021})}\BibitemShut {NoStop}%
\bibitem [{\citenamefont {Ran}\ \emph {et~al.}(2019)\citenamefont {Ran},
  \citenamefont {Eckberg}, \citenamefont {Ding}, \citenamefont {Furukawa},
  \citenamefont {Metz}, \citenamefont {Saha}, \citenamefont {Liu},
  \citenamefont {Zic}, \citenamefont {Kim}, \citenamefont {Paglione},\ and\
  \citenamefont {Butch}}]{Ran2019}%
  \BibitemOpen
  \bibfield  {author} {\bibinfo {author} {\bibfnamefont {S.}~\bibnamefont
  {Ran}}, \bibinfo {author} {\bibfnamefont {C.}~\bibnamefont {Eckberg}},
  \bibinfo {author} {\bibfnamefont {Q.~P.}\ \bibnamefont {Ding}}, \bibinfo
  {author} {\bibfnamefont {Y.}~\bibnamefont {Furukawa}}, \bibinfo {author}
  {\bibfnamefont {T.}~\bibnamefont {Metz}}, \bibinfo {author} {\bibfnamefont
  {S.~R.}\ \bibnamefont {Saha}}, \bibinfo {author} {\bibfnamefont {I.~L.}\
  \bibnamefont {Liu}}, \bibinfo {author} {\bibfnamefont {M.}~\bibnamefont
  {Zic}}, \bibinfo {author} {\bibfnamefont {H.}~\bibnamefont {Kim}}, \bibinfo
  {author} {\bibfnamefont {J.}~\bibnamefont {Paglione}}, \ and\ \bibinfo
  {author} {\bibfnamefont {N.~P.}\ \bibnamefont {Butch}},\ }\href {\doibase
  10.1126/science.aav8645} {\bibfield  {journal} {\bibinfo  {journal}
  {Science}\ }\textbf {\bibinfo {volume} {365}},\ \bibinfo {pages} {684}
  (\bibinfo {year} {2019})}\BibitemShut {NoStop}%
\bibitem [{\citenamefont {Jiao}\ \emph {et~al.}(2020)\citenamefont {Jiao},
  \citenamefont {Howard}, \citenamefont {Ran}, \citenamefont {Wang},
  \citenamefont {Rodriguez}, \citenamefont {Sigrist}, \citenamefont {Wang},
  \citenamefont {Butch},\ and\ \citenamefont {Madhavan}}]{Jiao2020chiral}%
  \BibitemOpen
  \bibfield  {author} {\bibinfo {author} {\bibfnamefont {L.}~\bibnamefont
  {Jiao}}, \bibinfo {author} {\bibfnamefont {S.}~\bibnamefont {Howard}},
  \bibinfo {author} {\bibfnamefont {S.}~\bibnamefont {Ran}}, \bibinfo {author}
  {\bibfnamefont {Z.}~\bibnamefont {Wang}}, \bibinfo {author} {\bibfnamefont
  {J.~O.}\ \bibnamefont {Rodriguez}}, \bibinfo {author} {\bibfnamefont
  {M.}~\bibnamefont {Sigrist}}, \bibinfo {author} {\bibfnamefont
  {Z.}~\bibnamefont {Wang}}, \bibinfo {author} {\bibfnamefont {N.~P.}\
  \bibnamefont {Butch}}, \ and\ \bibinfo {author} {\bibfnamefont
  {V.}~\bibnamefont {Madhavan}},\ }\href {\doibase 10.1038/s41586-020-2122-2}
  {\bibfield  {journal} {\bibinfo  {journal} {Nature}\ }\textbf {\bibinfo
  {volume} {579}},\ \bibinfo {pages} {523} (\bibinfo {year}
  {2020})}\BibitemShut {NoStop}%
\bibitem [{\citenamefont {Yanase}(2014)}]{Yanase2014}%
  \BibitemOpen
  \bibfield  {author} {\bibinfo {author} {\bibfnamefont {Y.}~\bibnamefont
  {Yanase}},\ }\href {\doibase 10.7566/JPSJ.88.103701} {\bibfield  {journal}
  {\bibinfo  {journal} {J. Phys. Soc. Jpn.}\ }\textbf {\bibinfo {volume}
  {83}},\ \bibinfo {pages} {014703} (\bibinfo {year} {2014})}\BibitemShut
  {NoStop}%
\bibitem [{\citenamefont {Hayami}\ \emph {et~al.}(2014)\citenamefont {Hayami},
  \citenamefont {Kusunose},\ and\ \citenamefont {Motome}}]{Hayami2014}%
  \BibitemOpen
  \bibfield  {author} {\bibinfo {author} {\bibfnamefont {S.}~\bibnamefont
  {Hayami}}, \bibinfo {author} {\bibfnamefont {H.}~\bibnamefont {Kusunose}}, \
  and\ \bibinfo {author} {\bibfnamefont {Y.}~\bibnamefont {Motome}},\ }\href
  {\doibase 10.1103/PhysRevB.90.024432} {\bibfield  {journal} {\bibinfo
  {journal} {Phys. Rev. B}\ }\textbf {\bibinfo {volume} {90}},\ \bibinfo
  {pages} {024432} (\bibinfo {year} {2014})}\BibitemShut {NoStop}%
\bibitem [{\citenamefont {Rikken}\ \emph {et~al.}(2001)\citenamefont {Rikken},
  \citenamefont {F\"olling},\ and\ \citenamefont {Wyder}}]{Rikken2001-il}%
  \BibitemOpen
  \bibfield  {author} {\bibinfo {author} {\bibfnamefont {G.~L. J.~A.}\
  \bibnamefont {Rikken}}, \bibinfo {author} {\bibfnamefont {J.}~\bibnamefont
  {F\"olling}}, \ and\ \bibinfo {author} {\bibfnamefont {P.}~\bibnamefont
  {Wyder}},\ }\href {\doibase 10.1103/PhysRevLett.87.236602} {\bibfield
  {journal} {\bibinfo  {journal} {Phys. Rev. Lett.}\ }\textbf {\bibinfo
  {volume} {87}},\ \bibinfo {pages} {236602} (\bibinfo {year}
  {2001})}\BibitemShut {NoStop}%
\bibitem [{\citenamefont {Watanabe}\ and\ \citenamefont
  {Yanase}(2017)}]{HW2017_magnetopiezo}%
  \BibitemOpen
  \bibfield  {author} {\bibinfo {author} {\bibfnamefont {H.}~\bibnamefont
  {Watanabe}}\ and\ \bibinfo {author} {\bibfnamefont {Y.}~\bibnamefont
  {Yanase}},\ }\href {\doibase 10.1103/PhysRevB.96.064432} {\bibfield
  {journal} {\bibinfo  {journal} {Phys. Rev. B}\ }\textbf {\bibinfo {volume}
  {96}},\ \bibinfo {pages} {064432} (\bibinfo {year} {2017})}\BibitemShut
  {NoStop}%
\bibitem [{\citenamefont {Shiomi}\ \emph {et~al.}(2019)\citenamefont {Shiomi},
  \citenamefont {Watanabe}, \citenamefont {Masuda}, \citenamefont {Takahashi},
  \citenamefont {Yanase},\ and\ \citenamefont {Ishiwata}}]{Shiomi2019}%
  \BibitemOpen
  \bibfield  {author} {\bibinfo {author} {\bibfnamefont {Y.}~\bibnamefont
  {Shiomi}}, \bibinfo {author} {\bibfnamefont {H.}~\bibnamefont {Watanabe}},
  \bibinfo {author} {\bibfnamefont {H.}~\bibnamefont {Masuda}}, \bibinfo
  {author} {\bibfnamefont {H.}~\bibnamefont {Takahashi}}, \bibinfo {author}
  {\bibfnamefont {Y.}~\bibnamefont {Yanase}}, \ and\ \bibinfo {author}
  {\bibfnamefont {S.}~\bibnamefont {Ishiwata}},\ }\href {\doibase
  10.1103/PhysRevLett.122.127207} {\bibfield  {journal} {\bibinfo  {journal}
  {Phys. Rev. Lett.}\ }\textbf {\bibinfo {volume} {122}},\ \bibinfo {pages}
  {127207} (\bibinfo {year} {2019})}\BibitemShut {NoStop}%
\bibitem [{\citenamefont {Ramires}\ and\ \citenamefont
  {Sigrist}(2016)}]{SCF2016}%
  \BibitemOpen
  \bibfield  {author} {\bibinfo {author} {\bibfnamefont {A.}~\bibnamefont
  {Ramires}}\ and\ \bibinfo {author} {\bibfnamefont {M.}~\bibnamefont
  {Sigrist}},\ }\href {\doibase 10.1103/PhysRevB.94.104501} {\bibfield
  {journal} {\bibinfo  {journal} {Phys. Rev. B}\ }\textbf {\bibinfo {volume}
  {94}},\ \bibinfo {pages} {104501} (\bibinfo {year} {2016})}\BibitemShut
  {NoStop}%
\bibitem [{\citenamefont {Ramires}\ \emph {et~al.}(2018)\citenamefont
  {Ramires}, \citenamefont {Agterberg},\ and\ \citenamefont
  {Sigrist}}]{SCF2018}%
  \BibitemOpen
  \bibfield  {author} {\bibinfo {author} {\bibfnamefont {A.}~\bibnamefont
  {Ramires}}, \bibinfo {author} {\bibfnamefont {D.~F.}\ \bibnamefont
  {Agterberg}}, \ and\ \bibinfo {author} {\bibfnamefont {M.}~\bibnamefont
  {Sigrist}},\ }\href {\doibase 10.1103/PhysRevB.98.024501} {\bibfield
  {journal} {\bibinfo  {journal} {Phys. Rev. B}\ }\textbf {\bibinfo {volume}
  {98}},\ \bibinfo {pages} {024501} (\bibinfo {year} {2018})}\BibitemShut
  {NoStop}%
\bibitem [{\citenamefont {Aoki}\ \emph {et~al.}(2019)\citenamefont {Aoki},
  \citenamefont {Nakamura}, \citenamefont {Honda}, \citenamefont {Li},
  \citenamefont {Homma}, \citenamefont {Shimizu}, \citenamefont {Sato},
  \citenamefont {Knebel}, \citenamefont {Brison}, \citenamefont {Pourret},
  \citenamefont {Braithwaite}, \citenamefont {Lapertot}, \citenamefont {Niu},
  \citenamefont {Vali{\v{s}}ka}, \citenamefont {Harima},\ and\ \citenamefont
  {Flouquet}}]{Aoki2019}%
  \BibitemOpen
  \bibfield  {author} {\bibinfo {author} {\bibfnamefont {D.}~\bibnamefont
  {Aoki}}, \bibinfo {author} {\bibfnamefont {A.}~\bibnamefont {Nakamura}},
  \bibinfo {author} {\bibfnamefont {F.}~\bibnamefont {Honda}}, \bibinfo
  {author} {\bibfnamefont {D.~X.}\ \bibnamefont {Li}}, \bibinfo {author}
  {\bibfnamefont {Y.}~\bibnamefont {Homma}}, \bibinfo {author} {\bibfnamefont
  {Y.}~\bibnamefont {Shimizu}}, \bibinfo {author} {\bibfnamefont {Y.~J.}\
  \bibnamefont {Sato}}, \bibinfo {author} {\bibfnamefont {G.}~\bibnamefont
  {Knebel}}, \bibinfo {author} {\bibfnamefont {J.~P.}\ \bibnamefont {Brison}},
  \bibinfo {author} {\bibfnamefont {A.}~\bibnamefont {Pourret}}, \bibinfo
  {author} {\bibfnamefont {D.}~\bibnamefont {Braithwaite}}, \bibinfo {author}
  {\bibfnamefont {G.}~\bibnamefont {Lapertot}}, \bibinfo {author}
  {\bibfnamefont {Q.}~\bibnamefont {Niu}}, \bibinfo {author} {\bibfnamefont
  {M.}~\bibnamefont {Vali{\v{s}}ka}}, \bibinfo {author} {\bibfnamefont
  {H.}~\bibnamefont {Harima}}, \ and\ \bibinfo {author} {\bibfnamefont
  {J.}~\bibnamefont {Flouquet}},\ }\href {\doibase 10.7566/JPSJ.88.043702}
  {\bibfield  {journal} {\bibinfo  {journal} {J. Phys. Soc. Jpn.}\ }\textbf
  {\bibinfo {volume} {88}},\ \bibinfo {pages} {043702} (\bibinfo {year}
  {2019})}\BibitemShut {NoStop}%
\bibitem [{\citenamefont {Knafo}\ \emph {et~al.}(2019)\citenamefont {Knafo},
  \citenamefont {Vali{\v{s}}ka}, \citenamefont {Braithwaite}, \citenamefont
  {Lapertot}, \citenamefont {Knebel}, \citenamefont {Pourret}, \citenamefont
  {Brison}, \citenamefont {Flouquet},\ and\ \citenamefont
  {Aoki}}]{Knafo2019RSC}%
  \BibitemOpen
  \bibfield  {author} {\bibinfo {author} {\bibfnamefont {W.}~\bibnamefont
  {Knafo}}, \bibinfo {author} {\bibfnamefont {M.}~\bibnamefont
  {Vali{\v{s}}ka}}, \bibinfo {author} {\bibfnamefont {D.}~\bibnamefont
  {Braithwaite}}, \bibinfo {author} {\bibfnamefont {G.}~\bibnamefont
  {Lapertot}}, \bibinfo {author} {\bibfnamefont {G.}~\bibnamefont {Knebel}},
  \bibinfo {author} {\bibfnamefont {A.}~\bibnamefont {Pourret}}, \bibinfo
  {author} {\bibfnamefont {J.~P.}\ \bibnamefont {Brison}}, \bibinfo {author}
  {\bibfnamefont {J.}~\bibnamefont {Flouquet}}, \ and\ \bibinfo {author}
  {\bibfnamefont {D.}~\bibnamefont {Aoki}},\ }\href {\doibase
  10.7566/JPSJ.88.063705} {\bibfield  {journal} {\bibinfo  {journal} {J. Phys.
  Soc. Jpn.}\ }\textbf {\bibinfo {volume} {88}},\ \bibinfo {pages} {063705}
  (\bibinfo {year} {2019})}\BibitemShut {NoStop}%
\bibitem [{\citenamefont {Metz}\ \emph {et~al.}(2019)\citenamefont {Metz},
  \citenamefont {Bae}, \citenamefont {Ran}, \citenamefont {Liu}, \citenamefont
  {Eo}, \citenamefont {Fuhrman}, \citenamefont {Agterberg}, \citenamefont
  {Anlage}, \citenamefont {Butch},\ and\ \citenamefont
  {Paglione}}]{Metz2019thermal}%
  \BibitemOpen
  \bibfield  {author} {\bibinfo {author} {\bibfnamefont {T.}~\bibnamefont
  {Metz}}, \bibinfo {author} {\bibfnamefont {S.}~\bibnamefont {Bae}}, \bibinfo
  {author} {\bibfnamefont {S.}~\bibnamefont {Ran}}, \bibinfo {author}
  {\bibfnamefont {I.~L.}\ \bibnamefont {Liu}}, \bibinfo {author} {\bibfnamefont
  {Y.~S.}\ \bibnamefont {Eo}}, \bibinfo {author} {\bibfnamefont {W.~T.}\
  \bibnamefont {Fuhrman}}, \bibinfo {author} {\bibfnamefont {D.~F.}\
  \bibnamefont {Agterberg}}, \bibinfo {author} {\bibfnamefont {S.~M.}\
  \bibnamefont {Anlage}}, \bibinfo {author} {\bibfnamefont {N.~P.}\
  \bibnamefont {Butch}}, \ and\ \bibinfo {author} {\bibfnamefont
  {J.}~\bibnamefont {Paglione}},\ }\href {\doibase 10.1103/PhysRevB.100.220504}
  {\bibfield  {journal} {\bibinfo  {journal} {Phys. Rev. B}\ }\textbf {\bibinfo
  {volume} {100}},\ \bibinfo {pages} {220504(R)} (\bibinfo {year}
  {2019})}\BibitemShut {NoStop}%
\bibitem [{\citenamefont {Nakamine}\ \emph {et~al.}(2021)\citenamefont
  {Nakamine}, \citenamefont {Kinjo}, \citenamefont {Kitagawa}, \citenamefont
  {Ishida}, \citenamefont {Tokunaga}, \citenamefont {Sakai}, \citenamefont
  {Kambe}, \citenamefont {Nakamura}, \citenamefont {Shimizu}, \citenamefont
  {Homma}, \citenamefont {Li}, \citenamefont {Honda},\ and\ \citenamefont
  {Aoki}}]{Nakamine2021}%
  \BibitemOpen
  \bibfield  {author} {\bibinfo {author} {\bibfnamefont {G.}~\bibnamefont
  {Nakamine}}, \bibinfo {author} {\bibfnamefont {K.}~\bibnamefont {Kinjo}},
  \bibinfo {author} {\bibfnamefont {S.}~\bibnamefont {Kitagawa}}, \bibinfo
  {author} {\bibfnamefont {K.}~\bibnamefont {Ishida}}, \bibinfo {author}
  {\bibfnamefont {Y.}~\bibnamefont {Tokunaga}}, \bibinfo {author}
  {\bibfnamefont {H.}~\bibnamefont {Sakai}}, \bibinfo {author} {\bibfnamefont
  {S.}~\bibnamefont {Kambe}}, \bibinfo {author} {\bibfnamefont
  {A.}~\bibnamefont {Nakamura}}, \bibinfo {author} {\bibfnamefont
  {Y.}~\bibnamefont {Shimizu}}, \bibinfo {author} {\bibfnamefont
  {Y.}~\bibnamefont {Homma}}, \bibinfo {author} {\bibfnamefont
  {D.}~\bibnamefont {Li}}, \bibinfo {author} {\bibfnamefont {F.}~\bibnamefont
  {Honda}}, \ and\ \bibinfo {author} {\bibfnamefont {D.}~\bibnamefont {Aoki}},\
  }\href {\doibase 10.1103/PhysRevB.103.L100503} {\bibfield  {journal}
  {\bibinfo  {journal} {Phys. Rev. B}\ }\textbf {\bibinfo {volume} {103}},\
  \bibinfo {pages} {L100503} (\bibinfo {year} {2021})}\BibitemShut {NoStop}%
\bibitem [{\citenamefont {Thomas}\ \emph {et~al.}(2020)\citenamefont {Thomas},
  \citenamefont {Santos}, \citenamefont {Christensen}, \citenamefont {Asaba},
  \citenamefont {Ronning}, \citenamefont {Thompson}, \citenamefont {Bauer},
  \citenamefont {Fernandes}, \citenamefont {Fabbris},\ and\ \citenamefont
  {Rosa}}]{Thomas2020}%
  \BibitemOpen
  \bibfield  {author} {\bibinfo {author} {\bibfnamefont {S.~M.}\ \bibnamefont
  {Thomas}}, \bibinfo {author} {\bibfnamefont {F.~B.}\ \bibnamefont {Santos}},
  \bibinfo {author} {\bibfnamefont {M.~H.}\ \bibnamefont {Christensen}},
  \bibinfo {author} {\bibfnamefont {T.}~\bibnamefont {Asaba}}, \bibinfo
  {author} {\bibfnamefont {F.}~\bibnamefont {Ronning}}, \bibinfo {author}
  {\bibfnamefont {J.~D.}\ \bibnamefont {Thompson}}, \bibinfo {author}
  {\bibfnamefont {E.~D.}\ \bibnamefont {Bauer}}, \bibinfo {author}
  {\bibfnamefont {R.~M.}\ \bibnamefont {Fernandes}}, \bibinfo {author}
  {\bibfnamefont {G.}~\bibnamefont {Fabbris}}, \ and\ \bibinfo {author}
  {\bibfnamefont {P.~F.}\ \bibnamefont {Rosa}},\ }\href {\doibase
  10.1126/sciadv.abc8709} {\bibfield  {journal} {\bibinfo  {journal} {Sci.
  Adv.}\ }\textbf {\bibinfo {volume} {6}},\ \bibinfo {pages} {eabc8709}
  (\bibinfo {year} {2020})}\BibitemShut {NoStop}%
\bibitem [{\citenamefont {Ishizuka}\ \emph {et~al.}(2019)\citenamefont
  {Ishizuka}, \citenamefont {Sumita}, \citenamefont {Daido},\ and\
  \citenamefont {Yanase}}]{Ishizuka2019}%
  \BibitemOpen
  \bibfield  {author} {\bibinfo {author} {\bibfnamefont {J.}~\bibnamefont
  {Ishizuka}}, \bibinfo {author} {\bibfnamefont {S.}~\bibnamefont {Sumita}},
  \bibinfo {author} {\bibfnamefont {A.}~\bibnamefont {Daido}}, \ and\ \bibinfo
  {author} {\bibfnamefont {Y.}~\bibnamefont {Yanase}},\ }\href {\doibase
  10.1103/PhysRevLett.123.217001} {\bibfield  {journal} {\bibinfo  {journal}
  {Phys. Rev. Lett.}\ }\textbf {\bibinfo {volume} {123}},\ \bibinfo {pages}
  {217001} (\bibinfo {year} {2019})}\BibitemShut {NoStop}%
\bibitem [{\citenamefont {Hiranuma}\ and\ \citenamefont
  {Fujimoto}(2021)}]{Hiranuma2021}%
  \BibitemOpen
  \bibfield  {author} {\bibinfo {author} {\bibfnamefont {K.}~\bibnamefont
  {Hiranuma}}\ and\ \bibinfo {author} {\bibfnamefont {S.}~\bibnamefont
  {Fujimoto}},\ }\href {\doibase 10.7566/JPSJ.90.034707} {\bibfield  {journal}
  {\bibinfo  {journal} {J. Phys. Soc. Jpn.}\ }\textbf {\bibinfo {volume}
  {90}},\ \bibinfo {pages} {034707} (\bibinfo {year} {2021})}\BibitemShut
  {NoStop}%
\bibitem [{\citenamefont {Xu}\ \emph {et~al.}(2019)\citenamefont {Xu},
  \citenamefont {Sheng},\ and\ \citenamefont {Yang}}]{Xu2019DMFT}%
  \BibitemOpen
  \bibfield  {author} {\bibinfo {author} {\bibfnamefont {Y.}~\bibnamefont
  {Xu}}, \bibinfo {author} {\bibfnamefont {Y.}~\bibnamefont {Sheng}}, \ and\
  \bibinfo {author} {\bibfnamefont {Y.-f.}\ \bibnamefont {Yang}},\ }\href
  {\doibase 10.1103/PhysRevLett.123.217002} {\bibfield  {journal} {\bibinfo
  {journal} {Phys. Rev. Lett.}\ }\textbf {\bibinfo {volume} {123}},\ \bibinfo
  {pages} {217002} (\bibinfo {year} {2019})}\BibitemShut {NoStop}%
\bibitem [{\citenamefont {Shishidou}\ \emph {et~al.}(2021)\citenamefont
  {Shishidou}, \citenamefont {Suh}, \citenamefont {Brydon}, \citenamefont
  {Weinert},\ and\ \citenamefont {Agterberg}}]{Shishidou2021}%
  \BibitemOpen
  \bibfield  {author} {\bibinfo {author} {\bibfnamefont {T.}~\bibnamefont
  {Shishidou}}, \bibinfo {author} {\bibfnamefont {H.~G.}\ \bibnamefont {Suh}},
  \bibinfo {author} {\bibfnamefont {P.~M.~R.}\ \bibnamefont {Brydon}}, \bibinfo
  {author} {\bibfnamefont {M.}~\bibnamefont {Weinert}}, \ and\ \bibinfo
  {author} {\bibfnamefont {D.~F.}\ \bibnamefont {Agterberg}},\ }\href {\doibase
  10.1103/PhysRevB.103.104504} {\bibfield  {journal} {\bibinfo  {journal}
  {Phys. Rev. B}\ }\textbf {\bibinfo {volume} {103}},\ \bibinfo {pages}
  {104504} (\bibinfo {year} {2021})}\BibitemShut {NoStop}%
\bibitem [{\citenamefont {Fischer}\ \emph {et~al.}(2011)\citenamefont
  {Fischer}, \citenamefont {Loder},\ and\ \citenamefont
  {Sigrist}}]{Fischer2011}%
  \BibitemOpen
  \bibfield  {author} {\bibinfo {author} {\bibfnamefont {M.~H.}\ \bibnamefont
  {Fischer}}, \bibinfo {author} {\bibfnamefont {F.}~\bibnamefont {Loder}}, \
  and\ \bibinfo {author} {\bibfnamefont {M.}~\bibnamefont {Sigrist}},\ }\href
  {\doibase 10.1103/PhysRevB.84.184533} {\bibfield  {journal} {\bibinfo
  {journal} {Phys. Rev. B}\ }\textbf {\bibinfo {volume} {84}},\ \bibinfo
  {pages} {184533} (\bibinfo {year} {2011})}\BibitemShut {NoStop}%
\bibitem [{\citenamefont {Wadley}\ \emph {et~al.}(2016)\citenamefont {Wadley},
  \citenamefont {Howells}, \citenamefont {Elezny}, \citenamefont {Andrews},
  \citenamefont {Hills}, \citenamefont {Campion}, \citenamefont {Novak},
  \citenamefont {Olejnik}, \citenamefont {Maccherozzi}, \citenamefont {Dhesi},
  \citenamefont {Martin}, \citenamefont {Wagner}, \citenamefont {Wunderlich},
  \citenamefont {Freimuth}, \citenamefont {Mokrousov}, \citenamefont {Kune},
  \citenamefont {Chauhan}, \citenamefont {Grzybowski}, \citenamefont
  {Rushforth}, \citenamefont {Edmonds}, \citenamefont {Gallagher},\ and\
  \citenamefont {Jungwirth}}]{Wadley2016}%
  \BibitemOpen
  \bibfield  {author} {\bibinfo {author} {\bibfnamefont {P.}~\bibnamefont
  {Wadley}}, \bibinfo {author} {\bibfnamefont {B.}~\bibnamefont {Howells}},
  \bibinfo {author} {\bibfnamefont {J.}~\bibnamefont {Elezny}}, \bibinfo
  {author} {\bibfnamefont {C.}~\bibnamefont {Andrews}}, \bibinfo {author}
  {\bibfnamefont {V.}~\bibnamefont {Hills}}, \bibinfo {author} {\bibfnamefont
  {R.~P.}\ \bibnamefont {Campion}}, \bibinfo {author} {\bibfnamefont
  {V.}~\bibnamefont {Novak}}, \bibinfo {author} {\bibfnamefont
  {K.}~\bibnamefont {Olejnik}}, \bibinfo {author} {\bibfnamefont
  {F.}~\bibnamefont {Maccherozzi}}, \bibinfo {author} {\bibfnamefont {S.~S.}\
  \bibnamefont {Dhesi}}, \bibinfo {author} {\bibfnamefont {S.~Y.}\ \bibnamefont
  {Martin}}, \bibinfo {author} {\bibfnamefont {T.}~\bibnamefont {Wagner}},
  \bibinfo {author} {\bibfnamefont {J.}~\bibnamefont {Wunderlich}}, \bibinfo
  {author} {\bibfnamefont {F.}~\bibnamefont {Freimuth}}, \bibinfo {author}
  {\bibfnamefont {Y.}~\bibnamefont {Mokrousov}}, \bibinfo {author}
  {\bibfnamefont {J.}~\bibnamefont {Kune}}, \bibinfo {author} {\bibfnamefont
  {J.~S.}\ \bibnamefont {Chauhan}}, \bibinfo {author} {\bibfnamefont {M.~J.}\
  \bibnamefont {Grzybowski}}, \bibinfo {author} {\bibfnamefont {A.~W.}\
  \bibnamefont {Rushforth}}, \bibinfo {author} {\bibfnamefont {K.~W.}\
  \bibnamefont {Edmonds}}, \bibinfo {author} {\bibfnamefont {B.~L.}\
  \bibnamefont {Gallagher}}, \ and\ \bibinfo {author} {\bibfnamefont
  {T.}~\bibnamefont {Jungwirth}},\ }\href {\doibase 10.1126/science.aab1031}
  {\bibfield  {journal} {\bibinfo  {journal} {Science}\ }\textbf {\bibinfo
  {volume} {351}},\ \bibinfo {pages} {587} (\bibinfo {year}
  {2016})}\BibitemShut {NoStop}%
\bibitem [{\citenamefont {Watanabe}\ and\ \citenamefont
  {Yanase}(2018)}]{Watanabe2018_DS}%
  \BibitemOpen
  \bibfield  {author} {\bibinfo {author} {\bibfnamefont {H.}~\bibnamefont
  {Watanabe}}\ and\ \bibinfo {author} {\bibfnamefont {Y.}~\bibnamefont
  {Yanase}},\ }\href {\doibase 10.1103/PhysRevB.98.220412} {\bibfield
  {journal} {\bibinfo  {journal} {Phys. Rev. B}\ }\textbf {\bibinfo {volume}
  {98}},\ \bibinfo {pages} {220412(R)} (\bibinfo {year} {2018})}\BibitemShut
  {NoStop}%
\end{thebibliography}

\begin{thebibliography}{3}%
\makeatletter
\providecommand \@ifxundefined [1]{%
 \@ifx{#1\undefined}
}%
\providecommand \@ifnum [1]{%
 \ifnum #1\expandafter \@firstoftwo
 \else \expandafter \@secondoftwo
 \fi
}%
\providecommand \@ifx [1]{%
 \ifx #1\expandafter \@firstoftwo
 \else \expandafter \@secondoftwo
 \fi
}%
\providecommand \natexlab [1]{#1}%
\providecommand \enquote  [1]{``#1''}%
\providecommand \bibnamefont  [1]{#1}%
\providecommand \bibfnamefont [1]{#1}%
\providecommand \citenamefont [1]{#1}%
\providecommand \href@noop [0]{\@secondoftwo}%
\providecommand \href [0]{\begingroup \@sanitize@url \@href}%
\providecommand \@href[1]{\@@startlink{#1}\@@href}%
\providecommand \@@href[1]{\endgroup#1\@@endlink}%
\providecommand \@sanitize@url [0]{\catcode `\\12\catcode `\$12\catcode
  `\&12\catcode `\#12\catcode `\^12\catcode `\_12\catcode `\%12\relax}%
\providecommand \@@startlink[1]{}%
\providecommand \@@endlink[0]{}%
\providecommand \url  [0]{\begingroup\@sanitize@url \@url }%
\providecommand \@url [1]{\endgroup\@href {#1}{\urlprefix }}%
\providecommand \urlprefix  [0]{URL }%
\providecommand \Eprint [0]{\href }%
\providecommand \doibase [0]{http://dx.doi.org/}%
\providecommand \selectlanguage [0]{\@gobble}%
\providecommand \bibinfo  [0]{\@secondoftwo}%
\providecommand \bibfield  [0]{\@secondoftwo}%
\providecommand \translation [1]{[#1]}%
\providecommand \BibitemOpen [0]{}%
\providecommand \bibitemStop [0]{}%
\providecommand \bibitemNoStop [0]{.\EOS\space}%
\providecommand \EOS [0]{\spacefactor3000\relax}%
\providecommand \BibitemShut  [1]{\csname bibitem#1\endcsname}%
\let\auto@bib@innerbib\@empty
\bibitem [{\citenamefont {Brydon}\ \emph {et~al.}(2019)\citenamefont {Brydon},
  \citenamefont {Abergel}, \citenamefont {Agterberg},\ and\ \citenamefont
  {Yakovenko}}]{S_Brydon2019AHE}%
  \BibitemOpen
  \bibfield  {author} {\bibinfo {author} {\bibfnamefont {P.~M.~R.}\
  \bibnamefont {Brydon}}, \bibinfo {author} {\bibfnamefont {D.~S.~L.}\
  \bibnamefont {Abergel}}, \bibinfo {author} {\bibfnamefont {D.~F.}\
  \bibnamefont {Agterberg}}, \ and\ \bibinfo {author} {\bibfnamefont {V.~M.}\
  \bibnamefont {Yakovenko}},\ }\href {\doibase 10.1103/PhysRevX.9.031025}
  {\bibfield  {journal} {\bibinfo  {journal} {Phys. Rev. X}\ }\textbf {\bibinfo
  {volume} {9}},\ \bibinfo {pages} {031025} (\bibinfo {year}
  {2019})}\BibitemShut {NoStop}%
\bibitem [{\citenamefont {Denys}\ and\ \citenamefont
  {Brydon}(2021)}]{S_Brydon2021AHE}%
  \BibitemOpen
  \bibfield  {author} {\bibinfo {author} {\bibfnamefont {M.~D.~E.}\
  \bibnamefont {Denys}}\ and\ \bibinfo {author} {\bibfnamefont {P.~M.~R.}\
  \bibnamefont {Brydon}},\ }\href {\doibase 10.1103/PhysRevB.103.094503}
  {\bibfield  {journal} {\bibinfo  {journal} {Phys. Rev. B}\ }\textbf {\bibinfo
  {volume} {103}},\ \bibinfo {pages} {094503} (\bibinfo {year}
  {2021})}\BibitemShut {NoStop}%
\bibitem [{\citenamefont {Ishizuka}\ and\ \citenamefont
  {Yanase}(2021)}]{S_Ishizuka2020}%
  \BibitemOpen
  \bibfield  {author} {\bibinfo {author} {\bibfnamefont {J.}~\bibnamefont
  {Ishizuka}}\ and\ \bibinfo {author} {\bibfnamefont {Y.}~\bibnamefont
  {Yanase}},\ }\href {\doibase 10.1103/PhysRevB.103.094504} {\bibfield
  {journal} {\bibinfo  {journal} {Phys. Rev. B}\ }\textbf {\bibinfo {volume}
  {103}},\ \bibinfo {pages} {094504} (\bibinfo {year} {2021})}\BibitemShut
  {NoStop}%
\end{thebibliography}
\providecommand{\noopsort}[1]{}\providecommand{\singleletter}[1]{#1}%
%

\end{document}